\documentclass[aps,pre,onecolumn,superscriptaddress]{revtex4}
\usepackage[dvips]{graphicx}
\usepackage[dvips]{color}
\usepackage{breakurl,amsmath,amssymb,pst-node,eepic}

\usepackage{subfigure}
\usepackage{tikz}
\usetikzlibrary{shapes, arrows, calc, positioning,matrix}
\usepackage{bbold}

\begin{document}

\title{A parametric approach to information filtering in complex networks: The P\'olya filter}

\author{Riccardo Marcaccioli}
\email{riccardo.marcaccioli.16@ucl.ac.uk}
\affiliation{Department of Computer Science, University College London, 66-72 Gower Street, London WC1E 6EA, UK}

\author{Giacomo Livan}
\email{g.livan@ucl.ac.uk}
\affiliation{Department of Computer Science, University College London, 66-72 Gower Street, London WC1E 6EA, UK}
\affiliation{Systemic Risk Centre, London School of Economics and Political Sciences, Houghton Street, London WC2A 2AE, UK}


\begin{abstract}
The increasing availability of data demands for techniques to filter information in large complex networks of interactions. A number of approaches have been proposed to extract network backbones by assessing the statistical significance of links against null hypotheses of random interaction. Yet, it is well known that the growth of most real-world networks is non-random, as past interactions between nodes typically increase the likelihood of further interaction. Here, we propose a  filtering methodology inspired by the P\'olya urn, a combinatorial model driven by a self-reinforcement mechanism, which relies on a family of null hypotheses that can be calibrated to assess which links are statistically significant with respect to a given network's own heterogeneity. We provide a full characterization of the filter, and show that it selects links based on a non-trivial interplay between their local importance and the importance of the nodes they belong to.
\end{abstract}

\maketitle

\section{Introduction}

A vast number of complex interacting systems can be represented as networks \cite{latora_totale2}. Over the last 20 years, Network Science has been successfully applied in a wide range of disciplines, from Biology to Finance and the Social Sciences \cite{totale_barabasi,totale_latora,totale_Newman,totale_Vespignani,econophysics_Mantegna}. One of the main reasons behind such a success is that oftentimes network representations of seemingly very diverse systems share a number of common characteristics. A recurrent feature of several natural and social networks is the lack of a typical scale \cite{totale_barabasi,Caldarelli_scalefree}, i.e., the marked heterogeneity of major structural features such as the degree or strength distributions.

Understanding which nodes and links represent a set of structurally relevant interactions can be of crucial importance to obtain parsimonious descriptions of complex networks, and, indeed, has contributed to shed light on the functioning of a variety of systems, ranging from biological \cite{biology_backbone,biology_backbone2}, social \cite{disease_backbone,social_movement_backbone}, financial \cite{finance_backbone} or even literature-related \cite{literature_backbone,literature_backbone2} systems. Furthermore, the size and, in some cases, the density of several real-world networks often prevent any meaningful visualization, and represent a major obstacle for clustering algorithms, which typically work well only with sparse systems \cite{community_detection,dbht}. Because of such challenges, a number of approaches to extract relevant information from complex networks have been developed over the years. Naturally, any filtering technique hinges on a definition of what type of information represents a signal as opposed to noise. As a result, the network backbones obtained through different filtering techniques carry different meanings and highlight different properties.

Early approaches to filtering focused on proximity networks, and relied on retaining interactions fulfilling some topological constraints. A seminal example of this kind of approach is the minimum spanning tree \cite{MST_mantegna}, which selects the tree with the highest total strength embedded in a network. Less constrained generalizations of such method are the planar maximally filtered graphs \cite{PMFG} and the triangulated maximally filtered graphs \cite{TMFG}, which reduce topological complexity by forcing the embedding of network backbones on a surface. 

Most of the methodologies initially proposed to filter information in weighted networks largely relied on discarding all links whose weights are below a certain global threshold \cite{global_TR1,global_TR2,global_TR3,global_TR4}, leading to backbones not reflecting the multiscale nature of the underlying network \cite{coupling_weight_topology}. This issue has been addressed by a different class of techniques, which resort to hypothesis testing in order to assess the statistical significance of each link in a network. The disparity filter \cite{backbone_vespignani}, which arguably represents one of most widely used filtering techniques, falls under this category, and relies on a null hypothesis of uniform distribution of a node's strength over its links. Such a method has been adopted as one of the main benchmarks against which the efficiency of  filtering techniques has been tested \cite{backbone_strana,backbone_battiston,gloss_filter,filtering_algo_Slater}. 

More recently, a procedure based on a null hypothesis of random connectivity (encoded as the urn problem described by the hypergeometric distribution) has been put forward \cite{bonferroni_filter,bonferroni_call,iori_mantegna}. Other recently proposed methodologies rely instead on frameworks inspired by Statistical Physics, where the properties of empirical networks are tested against those observed in an ensemble of null network models constrained to preserve, on average, the original networks' degree and strength sequences \cite{dianati2016unwinding,gemmetto2017irreducible}.

The above procedures provide top-down approaches based on well defined null hypotheses, against which all links in a network are tested individually. While this certainly presents advantages in terms of convenience, at the same time it can lead to a lack of flexibility, as different networks may display different levels of heterogeneity, to which a ``one-fits-all'' null hypothesis cannot adapt. Furthermore, most of the above filters are based on null hypotheses of partially random interactions. Yet, interactions in most natural and social systems are far from being random, as past activity naturally breeds further activity \cite{past_breeds_future1,past_breeds_future2}. 

Here, we propose a filtering methodology based on a null hypothesis designed to respond to the specific heterogeneity of a network. We shall do so through a statistical test based on the P\'olya urn, a well known combinatorial problem driven by a self-reinforcement mechanism according to which the observation of a certain event increases the probability of further observing it. Such a mechanism is governed by a single parameter $a$, which allows to tune the null hypothesis' tolerance to heterogeneity, and to study a continuous family of network backbones $\mathcal{P}_a$. In the following, we shall detail how the P\'olya filter works, both from an intuitive standpoint and by providing a full analytical characterization of the family of backbones it generates. In doing so, we shall show how the disparity filter can be recovered, with very good approximation, as a special case of the P\'olya filter for $a = 1$. We shall complement our analyses with two case studies to illustrate possible application of the P\'olya filter to real-world network data.

\section{Results}

\subsection*{The P\'olya Filter}

In the classic P\'olya urn problem, we are given an urn containing $B_0$ black balls and $R_0$ red balls. We randomly draw a ball from the urn, we observe its colour and put it back in the urn together with $a$ new balls of the same colour. When this process is repeated $n$ times, the probability of observing $x$ red balls follows the Beta-Binomial distribution \cite{Polya_book} with probability mass function $\mathbb{P} (x \mid n, \alpha, \beta) = \binom{n}{x} B(x+\alpha,n-x+\beta)/B(\alpha,\beta)$, where $B$ denotes the beta function and $\alpha = R_0 / a$, $\beta = B_0 / a$. In the following, we shall adapt this situation to a network setting. 

Let us denote the $N \times N$ symmetric adjacency matrix of an undirected weighted network with $N$ nodes as $W$. An entry $w_{ij} \in \mathbb{N}$ of such a matrix is the weight associated with the link connecting nodes $i$ and $j$, and $w_{ij} = w_{ji} = 0$ when there is no connection between $i$ and $j$. The degree $k_i = \sum_{j=1}^N \mathbb{1}(w_{ij})$ (where $\mathbb{1}$ denotes the indicator function) quantifies the number of connections between a node $i$ and other nodes in the network, while $s_i = \sum_{j=1}^N w_{ij}$ denotes the strength of a node $i$, which is a measure of its activity in the network. 

With the above notation, we can now rewrite the P\'olya urn problem in network terms. Assume we are interested in assessing the statistical significance of a certain weight $w$ falling on one of the links of a node with degree $k$ and total strength $s$. Following the above example, we can think of this as a drawing process from a P\'olya urn with $1$ red ball and $k-1$ black balls initially, where we want to measure the probability of drawing $w$ red balls in $s$ attempts. Such a probability reads
\begin{equation}\label{eq:Polya_filter}
\mathbb{P}(w \mid k,s,a ) = \binom{s}{w} \frac{B \left (\frac{1}{a} + w, \frac{k-1}{a} + s - w \right)}{B \left (\frac{1}{a}, \frac{k-1}{a} \right )} \ .
\end{equation}

\begin{figure}
\centering
\includegraphics[scale=0.4]{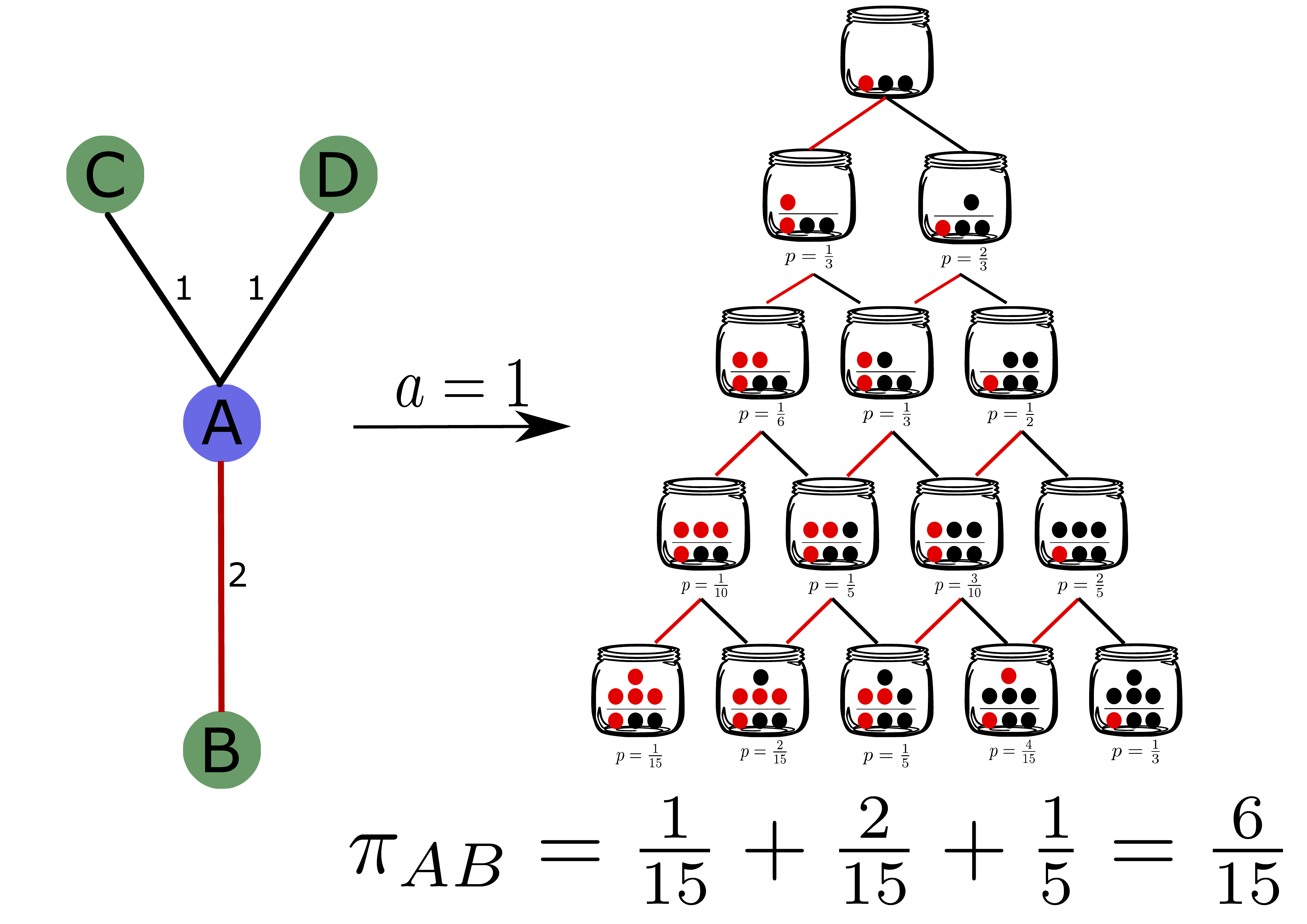}
\caption{Sketch of the P\'olya urn process in a network setting. In the toy example shown here, we aim to assess the statistical significance of the link of weight $w = 2$ (highlighted in red) connecting nodes $A$ and $B$, and we wish to do so from the viewpoint of node $A$, whose degree is $k=3$ and whose strength is $s = 4$. In the P\'olya urn analogy, this amounts to starting the urn with one red ball and $k-1=2$ black balls, and computing the probability of drawing at least $w$ red balls in $s$ draws (i.e., the probability that a node distributing its strength $s$ at random through a P\'olya process will assign a weight equal or larger than $w$ on the link under consideration). The right part of the Figure shows the possible configurations of the corresponding P\'olya urn (for $a=1$, which entails adding to the urn one ball of the same color of the latest ball drawn) over the $s$ draws, and their corresponding probabilities computed via Eq. \eqref{eq:Polya_filter}. The  $p$-value associated to the link is shown at the bottom of the Figure, and is computed as the sum over ``favourable'' outcomes (see Eq. \eqref{eq:Polya_prob_main}), i.e. urns containing at least $w = 2$ red balls (in addition to the one initially present in the urn) at the end of the process.}
\label{fig:Polya_representation}
\end{figure}

The above equation fully describes our class of null hypotheses. We shall assume that a node distributes the weights on its links following a P\'olya process whose reinforcement mechanism is governed by the parameter $a$. The rationale of such assumption lays in the flexibility introduced by such a parameter, which naturally captures situations where the more two nodes have interacted, the more further interactions between them become likely. In Fig. \ref{fig:Polya_representation} we provide a sketch of the P\'olya process adapted to a network setting.

Eq. \eqref{eq:Polya_filter} allows to assign a $p$-value to a link of weight $w$ as the sum over all possible ``favourable'' outcomes such that at least $w$ red balls have been drawn from the P\'olya urn after $s$ draws. This reads
\begin{equation} 
\label{eq:Polya_prob_main}
\pi_P (w \mid k,s,a) = 1- \sum_{x=0}^{w-1} \mathbb{P}(x \mid k,s,a ) \ ,
\end{equation}
and in Appendix \ref{sec:sn1} we provide an explicit formula for this quantity. Once the value of the free parameter $a$ has been set, two $p$-values can be assigned to the weight of each link in the network by applying Eq. \eqref{eq:Polya_prob_main} from the viewpoint of the two nodes it connects. The statistical significance of a weight is then assessed by comparing its associated $p$-values with a significance level. Since such a procedure involves testing all links in a network, it requires setting a univariate significance level $\alpha_u$ and applying a multiple hypothesis test correction. The two main options available in this respect are the Bonferroni \cite{bonferroni} and the false discovery rate (FDR) \cite{FDR} corrections. The benefits and limitations of the two methods have been largely debated \cite{criticone,critichino}, and choosing between them essentially boils down to the type of statistical error one is more inclined to accept. The Bonferroni correction is much stricter than the FDR and typically ensures very high precision, leading to a low probability of accepting false positives, at the cost of a potentially low accuracy, i.e., of rejecting true positives. Following \cite{bonferroni_filter}, in this work we shall adopt the Bonferroni correction: a link of weight $w$ will be validated and included in the P\'olya network backbone whenever at least one of its corresponding $p$-values will be such that $\pi_P < \alpha_u / L$, where $L$ is the number of statistical tests performed, which in the case of undirected network is given by twice the number of links in the network (in the case of a link between a node with degree $k=1$ and a node with $k>1$ we keep the link only if $\pi_P < \alpha_u / L$ for the node with degree greater than one.). 

We have introduced the P\'olya filter for weighted undirected networks but it can be easily extended to weighted directed networks (see Appendix \ref{sec:sn2}). In fact, the empirical analyses performed in the following are done on directed networks.

\subsection{The backbone family}

As mentioned above, the P\'olya filter generates a continuous family of network backbones $\mathcal{P}_a$, which we now seek to characterize as a function of the parameter $a$.

When $a=0$, the Beta-Binomial distribution (Eq. \eqref{eq:Polya_filter}) reduces to the Binomial distribution with parameters $s$ and $1/k$, i.e. $ \mathbb{P}(w \mid k,s,a=0) = \binom{s}{w} \left( \frac{1}{k}\right)^{w} \left(1-\frac{1}{k}\right)^{s-w}$. In the urn analogy, the $p$-value associated with a weight $w$ in this case corresponds to the probability of drawing at least $w$ red balls in $s$ attempts with simple replacement from an urn containing $1$ red balls and $k-1$ black balls.

When $a \to \infty$, instead, the P\'olya filter loses its dependency on the node strength $s$ and on the weight $w$. This corresponds to a situation where $a \gg k$ balls of the same color of the first drawn ball are added to the urn, and, as a result, all following extractions produce balls of the same color. Therefore, the probability of extracting at least $w$ red balls is the same of extracting one in the first draw, i.e., $1 / k$. This, in turn, leads to an empty network backbone, as the Bonferroni correction criterion cannot be met with such a probability.

\begin{figure}
\centering
\includegraphics[width=\linewidth]{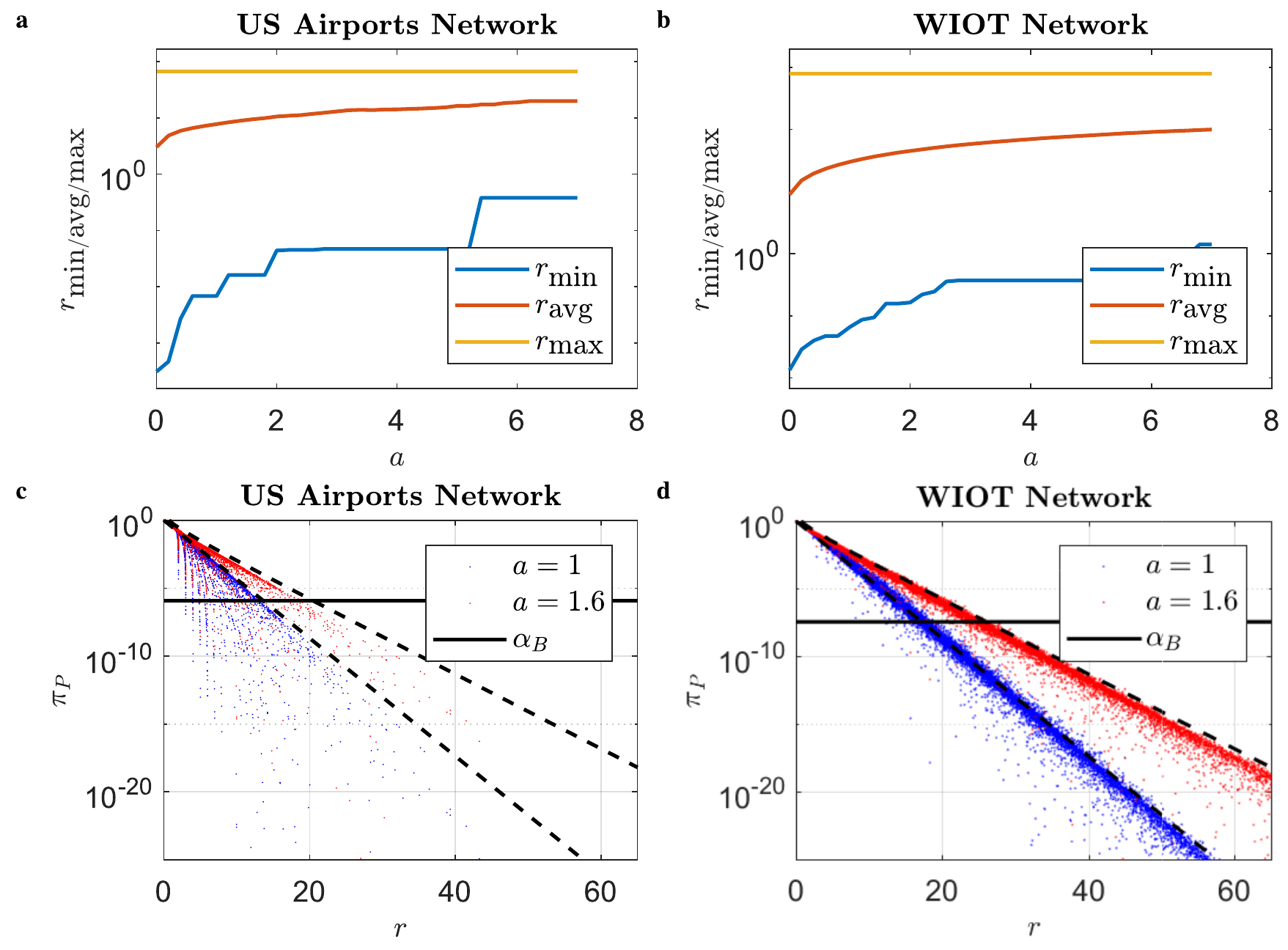}
\caption{Role of the parameter $r$ in the P\'olya backbone extraction process. (\textbf{a}) Evolution of the minimum, maximum, and average value of $r$ computed in P\'olya backbones for increasing values of $a$ with a univariate significance level $\alpha_U = 0.05$ in the US Airports network. (\textbf{b}) Same quantities computed in the WIOT network. (\textbf{c}) Scatter plots of the $p$-values associated with each link in the US Airports network against the corresponding value of the ratio $r$ for two different values of $a$ at a univariate significance level $\alpha_U = 0.05$. High values of $r$ are associated with $p$-values below the Bonferroni threshold $\alpha_B$ (solid black line), while the opposite is not always true. The black dashed lines illustrate the soft dependence on $r$ described by Eq. \eqref{eq:top_bound_p_Polya}. (\textbf{d}) Same plot for the WIOT network.}
\label{fig:ratio_explained_1}
\end{figure}

Between the two above limit cases, P\'olya network backbones monotonically shrink when the parameter $a$ is increased while keeping the statistical significance fixed, i.e.,
\begin{equation}\label{eq:included_sets}
w \in \mathcal{P}_{a_2} \quad \Rightarrow \quad w \in \mathcal{P}_{a_1} \quad \text{for} \ a_1 \leq a_2 \ .
\end{equation}
In other words, the largest P\'olya set is the one corresponding to $a=0$, and increasing $a$ progressively removes links from this set. This process is largely driven by a soft dependence of the P\'olya filter on the following ratio:
\begin{equation}\label{eq:r_def}
r = \frac{w}{s} \, k = \frac{w}{\langle w \rangle} \ ,
\end{equation}
where $\langle w \rangle = s/k$ is the average weight on the links of the node to which the link under analysis is attached. For any fixed value of the parameter $a$, the P\'olya filter tends to validate links associated with higher values of $r$. Moreover, higher values of $a$ lead to the progressive rejection of links with higher values of $r$, which in turn leads to the property in Eq. \eqref{eq:included_sets}. These results are illustrated in Fig.~\ref{fig:ratio_explained_1} on two network datasets (the 2017 US Airports network and the World Input-Output Database \cite{WIOT}, see Appendix \ref{sec:methods} for a brief description). Indeed, in the two bottom panels one can see that higher values of $r$ tend to be associated with a higher statistical significance (and that such significance, in turn, decreases as $a$ increases), although this is not a strict relationship and there are substantial exceptions. We show in Appendix \ref{sec:sn4} that these exceptions ensure that thresholding on $r$ does not give a backbone as topologically rich as the one obtained with the full P\'olya filter, and therefore the latter should be preferred. This dependence on $r$ is fully described in the Materials Section (see Eq. \eqref{eq:top_bound_p_Polya}), and is derived analytically in Appendix \ref{sec:sn3}.

In summary, the two quantities that drive the backbone extraction process are $a$ and $r$. First, the ratio $r$ couples a network's local topology (through the degree $k$) to the activity of nodes (through the strength $s$ and weight $w$) in a non-trivial way. The soft dependence of the P\'olya filter on such quantity is what ensures that its backbones retain the multiscale nature of the networks they are extracted from. The parameter $a$, instead, ensures the flexibility of the method thanks to the analytical control we have over it (see Appendix \ref{sec:sn5}), which can be exploited to tailor the backbone extraction process with respect to the network's own heterogeneity or other meaningful criteria. This will be showcased in the following Section. Moreover, let us mention that $a$ can be directly related to the statistical significance $\alpha$ used to assess the null hypothesis: the backbones generated by taking $a = a_1$ can approximately be considered equivalent to those associated with $a = a_2>a_1$, provided that a higher statistical significance is set. This is discussed in Appendix \ref{sec:methods} and numerical evidence for this is provided in Appendix \ref{sec:sn6}.

\subsection{Fixing the free parameter} 

The main benefit of the P\'olya filter is its flexibility, which allows to explore the network backbones obtained when setting different levels of tolerance to heterogeneity, as quantified by the parameter $a$. We devote this section to recommending possible criteria that would identify an optimal value of such a parameter. Clearly, the notion of optimality strongly depends on the specific application being considered. Therefore, we will recommend three different criteria.

\begin{itemize}
\item Sweeping: The P\'olya filter's monotonicity can be exploited to fix a desired level of sparsity of the resulting backbone with respect to the original network, and to identify the value of $a$ that achieves it. Namely, as a consequence of the property in Eq. \eqref{eq:included_sets}, the fraction of nodes, of edges, and of total strength retained in the P\'olya backbones are all monotonically non-increasing functions of the parameter $a$. Hence, starting from $a=0$, one can scan the backbone family $\mathcal{P}_a$ for increasing values of $a$ until a desired level of sparsity has been reached (e.g., $5\%$ of the nodes in the original network).
 
\item Maximum likelihood: Eq. \eqref{eq:Polya_filter} can be used to define a log-likelihood function, which can in turn be shown to have a maximum (see Appendix \ref{sec:sn5}). By definition, such a value corresponds to the P\'olya process whose self-reinforcement mechanism is the most likely to generate the network under study. Effectively, this amounts to identifying the value $a_{\textnormal{ML}}$ corresponding to the ``nullest'' model in the P\'olya family or, in other words, the P\'olya process that best captures the heterogeneity of the network under consideration. We further convey this point in Appendix \ref{sec:sn3} by showing on synthetic networks that the maximum likelihood estimates of the parameter $a$ are indeed sensitive to changes in the network's heterogeneity. As such, this criterion is particularly suited to applications where validating the backbone as a whole is a priority. As an example, we report here the values of $a_\mathrm{ML}$ of the two networks we study in this paper. We find $a_\mathrm{ML} = 4.5$ for the US Airports network and $a_\mathrm{ML} = 3.4$ for the WIOT network.

\item Salience: Lastly, we are going to propose an ad-hoc criterion based on a compromise between the information retained in a backbone and the information lost by filtering the network it is extracted from. We shall quantify the former in terms of salience \cite{salience}, a recently proposed yet well established measure of link importance, which can be loosely defined as the fraction of weighted shortest-path trees a link participates in. This is a non-local measure that has been shown to account for both the topological position of a link and for the magnitude of its associated weight (somewhat in analogy to the quantity in Eq. \eqref{eq:r_def}), and captures several essential transport properties. In Appendix \ref{sec:sn7} we show that, as $a$ increases, the links removed from P\'olya backbones are generally those with a lower salience. As a result, the average salience $\langle S(a) \rangle$ retained in the backbones $\mathcal{P}_a$ increases with $a$.\\
Measuring the quality of a backbone just in terms of average salience could lead, in most cases, to an excessive depletion of the network under study. This tendency can be contrasted by penalizing large differences between backbones and their original networks. We do so by introducing the two following optimality measures
\begin{equation} \label{eq:optimality}
O_1 = J(W,\mathcal{P}_a) \cdot \langle S(a) \rangle \;\; , \qquad  O_2 = F_n(a) \cdot \langle S(a) \rangle \ ,
\end{equation}
where we are weighting the average salience against the Jaccard similarity $J(W,\mathcal{P}_a)$ between the weights in the original network and those in the backbone, or against the fraction $F_n(a)$ of nodes retained in $\mathcal{P}_a$, respectively. Fig.~\ref{fig:optimal_Polya} shows the behavior of the above metrics as functions of $a$ in the two networks we study. As it can be seen, both metrics achieve a maximum $a^*$, which represents the optimal compromise between high salience and similarity with respect to the original network. 

\begin{figure}
\centering
\includegraphics[width=\linewidth]{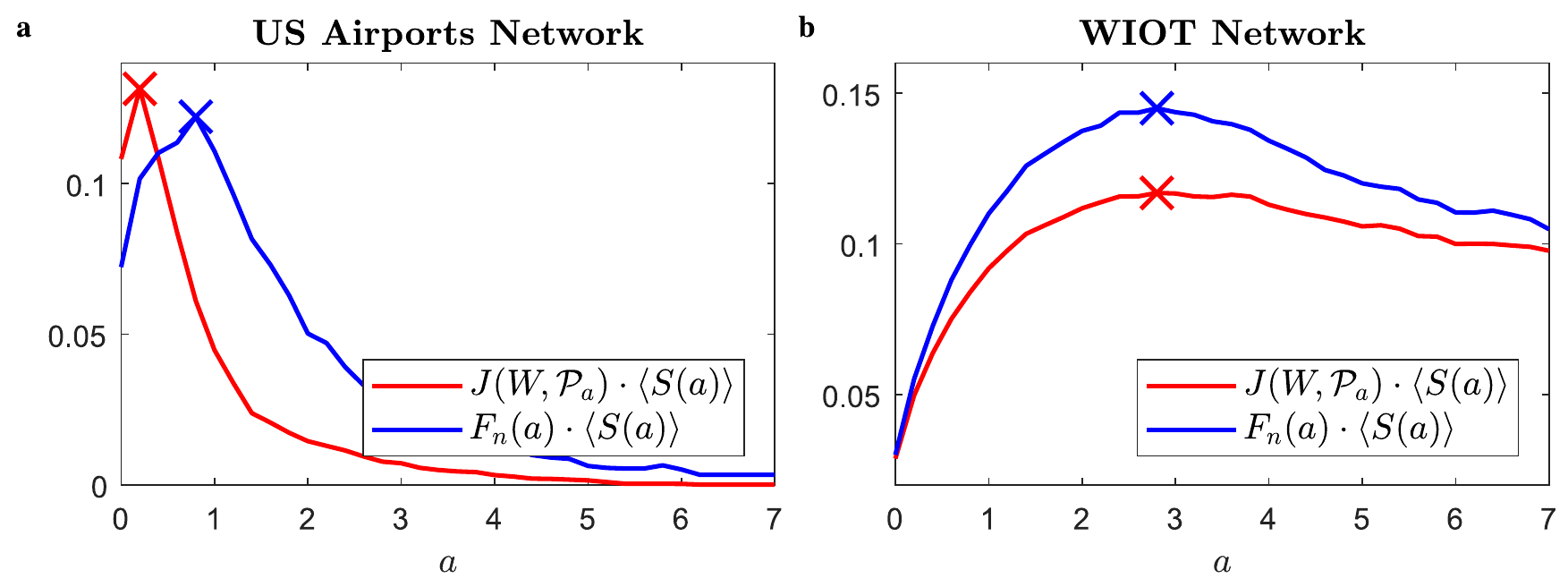}
\caption{Optimality measures $O_1$ and $O_2$. These are calculated on the extracted backbones (at a univariate significance level $\alpha_u = 0.05$) as a function of $a$. The optimal values are highlighted with a cross. (\textbf{a}) Optimality measures for the US Airports network. The optimal values are $a^* = 0.2$ for $O_1$ and $a^* = 0.8$ for $O_2$, respectively. (\textbf{b}) Same plot for the WIOT network. The optimal values are $a^* = 2.8$ for both $O_1$ and $O_2$.}
\label{fig:optimal_Polya}
\end{figure} 

\end{itemize}

\section{Comparisons with other network filters} 

In this Section and in Appendix \ref{sec:sn8} we further characterize the P\'olya filter's family of backbones through the comparison with some of the other available filtering techniques. In a nutshell, this will allow to show us that P\'olya backbones are typically sparse, salient and heterogeneous.

\begin{figure}
\centering
\includegraphics[width=0.95\linewidth]{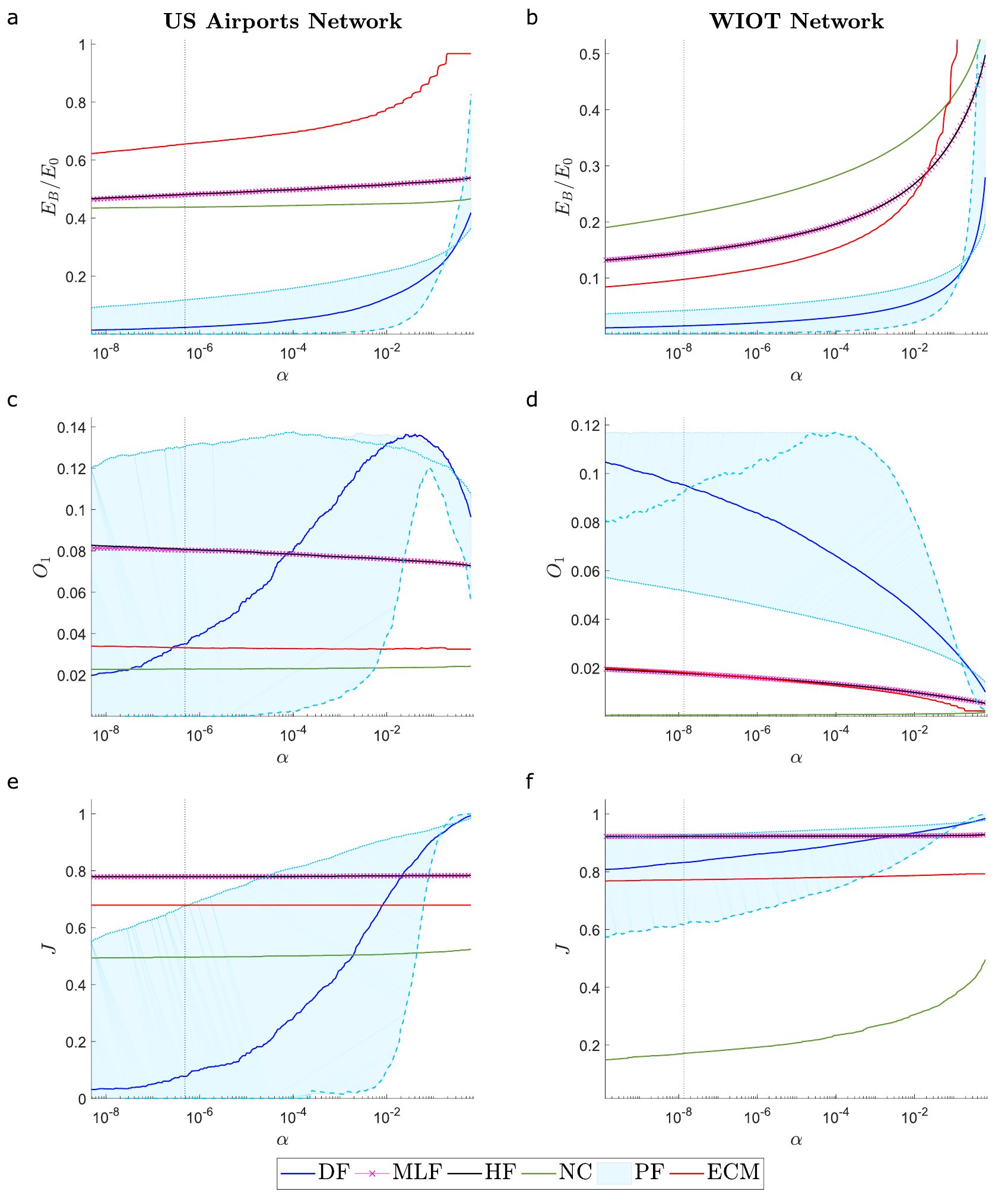}
\caption{Comparisons between the backbones generated by the P\'olya filter (PF) and other network filtering methods. The methods we consider are the Hypergeometric filter (HF), the Maximum-Likelihood filter (MLF), the Enhanced Configuration Model (ECM), the Noise-Corrected filter (NC), and the Disparity filer (DF), which corresponds to a large-strength approximation of the the P\'olya filter for $a=1$ (see the main text for references to the papers where those methods were introduced). All quantities are shown as a function of the multivariate significance level used in the tests. (\textbf{a})-(\textbf{b}) Fraction of links retained in the backbones with respect to the total number of links in the original networks. (\textbf{c})-(\textbf{d}) Value of the salience-related measure $O_1$ defined in Eq. \eqref{eq:optimality}. (\textbf{e})-(\textbf{f}) Jaccard similarity between the $B$ weights retained in the backbones and the top $B$ weights in the original networks. In all plots the light blue band correspond to all values measured in the P\'olya backbone family for $a \in [0.2, 7]$, with the light blue solid (dashed) line corresponding to $a=0.2$ ($a=7$); vertical dashed lines correspond to the Bonferroni-corrected $5\%$ significance level.}
\label{fig:comparisons}
\end{figure}

Fig. \ref{fig:comparisons} shows different properties of the P\'olya backbones of the US Airports and WIOT networks obtained for different multivariate significance levels $\alpha$ with those of the backbones obtained at the same statistical significance with the Hypergeometric Filter (HF) \cite{bonferroni_filter}, the Maximum-Likelihood filter (MLF) \cite{dianati2016unwinding}, the Enhanced Configuration Model (ECM) based on the canonical ensemble constrained both on degrees and strengths \cite{gemmetto2017irreducible}, the Noise-Corrected (NC) Bayesian filter proposed in \cite{coscia2017network}, and the Disparity Filter (DF) \cite{backbone_vespignani}, which in Appendix \ref{sec:methods} and in Appendix \ref{sec:sn3} we show to correspond to a large strength approximation of the P\'olya filter for $a=1$. Comparisons with the GloSS filter \cite{gloss_filter} were also performed, but their results are not reported due to the excessive sparsity of the backbones produced by such method when accounting for multiple hypothesis testing.

As it can be seen from the two upper panels (see also Appendix \ref{sec:sn8}), P\'olya backbones are considerably more parsimonious than those provided by the other filters considered. This is especially true when correcting for multiple hypothesis testing (the black vertical lines in each plot correspond to a Bonferroni-corrected univariate significance level of $0.05$, which is crucial to reduce the number of false positives retained in the backbones. In addition, when setting $a \simeq a_\mathrm{ML}$ (see previous Section), the P\'olya filter generates ultra-sparse backbones whose links are statistically significant with respect to the network's own heterogeneity. This will be further illustrated with a case study in the following Section.

The two middle panels show values of the optimality measure $O_1$ as a function of statistical significance. As it can be seen, for a wide range ot the parameter $a$ the P\'olya filter is able to strike a good balance between sparsity and salience, a property that is not shared by any other of the methods considered.

The two bottom panels demonstrate the heterogeneity of P\'olya backbones, by showing the Jaccard similarity between the $B$ weights retained in a backbone and the top $B$ weights in the original network. This essentially amounts to assessing how heterogeneous a network backbone is with respect to a ``naive'' backbone obtained simply by thresholding on weights. As one can see, the P\'olya filter generates backbones that are considerably more heterogeneous than those provided by the other methods, with the exception of the NC filter when applied to the WIOT network, where, however, such filter ends up discarding the more salient links.

The two bottom panels also show that the P\'olya filter is more responsive to statistical significance than the other methods. Indeed, P\'olya backbones are built around complex and sparse cores that correspond to links associated with very low $p$-values. As the threshold $\alpha$ increases, such cores are enriched by links with heavier weights which are structurally important for the network but classified as less statistically significant. Diversely, the other methods are much less responsive to $\alpha$, even when varied across several orders of magnitude.

The above properties are inherited by the disparity filter, which, as demonstrated in Appendix \ref{sec:methods} and in Appendix \ref{sec:sn3}, is a large-strength approximation of the P\'olya filter for $a=1$. In most cases (see also those in Appendix \ref{sec:sn8}), the disparity filter generates rather parsimonious backbones that are more salient and heterogeneous than most of the backbones produced by the other methods considered above. Yet, depending on the specific application or network, the disparity filter might be far from optimal within the P\'olya family. This is the case, for example, in the US Airports network, where the disparity filter backbone is rather sub-optimal in terms of salience, as demonstrated by the comparatively low value of $O_1$ it achieves within the P\'olya family.

All in all, the above results reiterate that the P\'olya filter's main advantage lies in its flexibility, which allows to tune the filter to the specific network or application under consideration. Moreover, the filter's ability to ``compress'' the salience and heterogeneity of the original networks in ultra-sparse backbones is unmatched by the other methods we considered. In the next Section we show how these properties can be exploited in order to gain insight on real-world networks.

The above observations can be largely replicated based on the additional comparisons shown in Appendix \ref{sec:sn8} between the above methods and the P\'olya filter.

\section{The short-haul backbone of the US Airports network} 

In the following we show how the P\'olya filter can be used to gather unique insights on the US Airports network. 

\begin{figure}
\centering
\includegraphics[width=\linewidth]{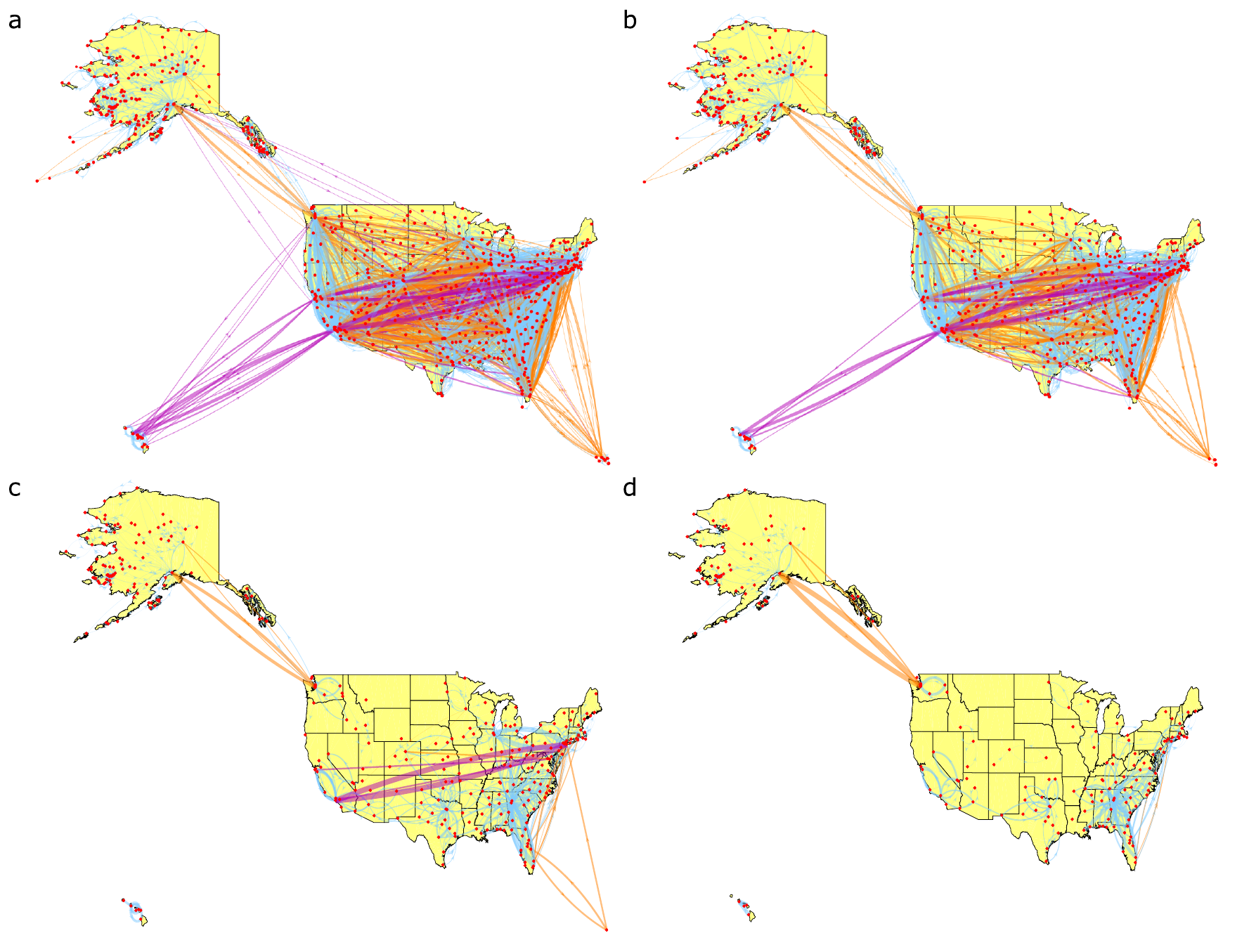}
\caption{P\'olya backbones of the US Airports network for different values of the filter's parameter $a$. (\textbf{a}) Backbone for $a=0.4$ (which is an intermediate value between the two that optimise the salience metrics in Eq. \eqref{eq:optimality}), where most long-haul flights between hubs are retained. (\textbf{b}) Backbone for $a=1$, approximately corresponding to the one obtained via the disparity filter. (\textbf{c}) Backbone for $a=2.6$, which is the highest value of the filter's parameter where a long-haul flight (New York - Los Angeles) is retained. (\textbf{d}) Backbone for $a=a_\mathrm{ML}=4.5$, where all long-haul flights and all connections between hubs have been filtered out.}
\label{fig:US_map}
\end{figure}

Fig. \ref{fig:US_map} shows the P\'olya filter's backbones of the US Airports network obtained for different values of the filter's parameter $a$. Thicker lines correspond to ``heavier'' links (i.e., routes with more passengers), while lines in blue, orange, and purple correspond, respectively, to short, medium, and long-haul flights according to the US Bureau of Transportation's classification.

As per Eq. \eqref{eq:included_sets}, higher values of $a$ lead to sparser backbones. The backbone in the top-left panel corresponds to $a=0.4$, (which is between the two values of $a$ that optimize the metrics defined in Eq. \eqref{eq:optimality}) is the most salient one. As such, it features the most crucial long-haul connections between hubs and/or the more geographically remote states (Alaska, Hawaii, and Puerto Rico). Most, although not all, of such connections are retained when setting $a=1$, which approximately corresponds to the disparity filter's backbone, shown in the top-right panel.

Things change considerably when increasing the filter's tolerance to heterogeneity through higher values of $a$. The backbone in the bottom-left panel is the one obtained for the highest value of $a$ that still allows to retain both connections between New York and Los Angeles ($a=2.6$), i.e., the two largest American cities. Notably, these are the only two long-haul connections remaining. Finally, when tuning the filter's tolerance to the network's own heterogeneity ($a = a_\mathrm{ML} = 4.5$), we obtain an ultra-sparse backbone, shown in the bottom right panel, where all long-haul flights and almost all connections between major cities and hubs have been filtered out. When projecting onto US states, this backbone is mostly made of two-way links between bordering or geographically close pairs of states. This is because long-haul connections are precisely those that determine the network's heterogeneity, while the links retained are those identified as statistically significant with respect to it. The only major hub still involved in a large number of connections is Atlanta, which is the busiest airport in the world and serves almost 20\% more passengers than the second busiest US airport. Notably, the links retained form a network of mostly regional and short-haul flights connecting airports that are often of secondary importance on the national scale.  Yet, these flights provide vital connections, carrying very large numbers of passengers relative to the overall heterogeneity of the broader transport system they are embedded in. This is well exemplified by Alaska, where a very large number of internal flights are validated.

\section{Predicting trade in the WIOT network} 

As an example of a practical use of our methodology, we show how the off-sample sample performance of a simple econometric model aimed at predicting trades in the WIOT network can be improved by using the P\'olya filter.

Understanding technological innovation ultimately hinges on the ability to foresee structural changes in the relationships between economic actors. Several studies have recently looked at this issue from a network perspective, where firms purchase goods from each other and combine them into more technologically sophisticated products (see, e.g., \cite{mcnerney2018production}). Within this framework, being able to predict changes in trading relationships can be of crucial importance in order to anticipate technological shifts and allow for an efficient allocation of investments.   

Here, we follow \cite{carvalho2014input,mcnerney2018production} and build a simple model to predict trading relationships in the WIOT dataset based on its network properties. We refer to Appendix \ref{sec:sn10} for a detailed description of the model. In short, it is a linear regression model aimed at predicting the future trading volume between two industrial sectors based on the relative importance of their past trading volume (with respect to their overall trading volume) and on their proximity in the network computed via the Leontief input-output matrix \cite{leontief1986input}.

We exploited such model to assess the potential benefits gained in terms of prediction accuracy when employing the P\'olya filter. Namely, we constructed P\'olya backbones of the annual WIOT networks from 2006 to 2010 both for $a=1$ (which essentially corresponds to the disparity filter) and for $a=a_\mathrm{ML}=3.2$. We used such backbones to calibrate the model (see Table \ref{tab:regression} in Appendix \ref{sec:sn10} for the model's coefficients and their significance) and to make out-of-sample predictions of the trading volumes of the links marked as significant in the three following years. We compared the predictive power of such models with that of the model calibrated on the full unfiltered WIOT network.

\begin{table}[!h]
\centering
  \caption{ $R^2$ coefficients of the model calibrated on the three different datasets when it is used to make out-of-sample predictions. }
  \label{tab:OS R2} 
{
\def\sym#1{\ifmmode^{#1}\else\(^{#1}\)\fi}
\begin{tabular}{l*{3}{c}}
\hline\hline\\[-1.8ex] 
& \multicolumn{3}{c}{\text{Out-of-sample $R^2$}} \\ 
\cline{2-4}\\[-1.8ex]  
&\multicolumn{1}{c}{2011}&\multicolumn{1}{c}{2012}&\multicolumn{1}{c}{2013}\\
\hline\\[-1.8ex] 
Unfiltered Networks        &      0.1349&      0.1371&      0.1367\\
[1em]
Backbones $\mathcal{P}_{a=1}$     &     0.1960&     0.1989&       0.1972\\
[1em]
Backbones $\mathcal{P}_{a_{ML}}$     &      0.2242&      0.2181&       0.2127\\
\hline\hline\\[-1.8ex] 
\end{tabular}
}
\end{table}

In Table \ref{tab:OS R2} we compare the predictive power of the model when calibrated on P\'olya backbones and on the full, unfiltered, WIOT network in terms of out-of-sample $R^2$ coefficients. As it can be seen, applying the P\'olya filter substantially improves the percentage of variance in the data explained by the model, with the best results being obtained when applying the filter for $a=a_\mathrm{ML}$.

These results further testify that the information contained in P\'olya backbones is substantial. Indeed, the full WIOT network contains $2.68 \times 10^6$ links, whereas the two P\'olya backbones employed above contain $4.89 \times 10^4$ and  $1.48 \times 10^4$ links for $a=1$ and $a=a_\mathrm{ML}$, respectively (see Table \ref{tab:regression} in Appendix \ref{sec:sn10}). This, in turn, means that the information lost by reducing the number of links by two orders of magnitude is more than offset by the higher overall informativeness of the networks generated by the filter.

\section{Discussion} 

In the era of Big Data, information filtering methods are needed more than ever to handle the dazzling complexity of both social and natural networked systems. In this paper, we have proposed a technique based on the P\'olya urn model to extract backbones of statistically relevant interactions between pairs of nodes in a network. In the network context, the parameter $a$ tuning the P\'olya model's self-reinforcement mechanism effectively becomes a tolerance to a network's heterogeneity. This, in turn, introduces an element of flexibility, which, to the best of our knowledge, other network filtering techniques do not provide.

Indeed, we have shown that the P\'olya filter generates a continuous family of network backbones. Depending on the specific application, the null hypothesis underpinning the filter can be chosen so as to have a different tolerance to heterogeneity. The low-tolerance regime ($a < 1$) corresponds to a rather loose filtering, suited to situations where the main goal is to filter out interactions that can be unquestionably identified as noise. On the other hand, the high-tolerance regime ($a > 1$) corresponds to increasingly restrictive tests, where only links of substantial structural importance survive. 

As we have shown, the link selection criterion underpinning the P\'olya filter is based on the interplay between topology and the local relative importance of a link, quantified by the parameter $r$. This, in turn, guarantees that the filter does not perform a naive link selection merely based on retaining high strength links connecting hubs, but instead ensures a non-trivial scanning of all the relevant scales of a network.

\appendix

\section{Methods} \label{sec:methods}

\subsection{Data}

In the following we provide a short description the datasets we employed to illustrate the P\'olya filter.
\begin{description}
\item[World Input Output Database]
The Database contains yearly aggregate economic transactions, measured in millions of dollars, between the industrial sectors of different countries from 2000 to 2014. The database features transactions between 64 sectors in 45 countries \cite{WIOT,WIOT2}. The resulting series of networks and their properties have been analyzed extensively in a number of studies \cite{WIOT_network,WIOT_stenley,WIOT_employment}. The dataset we are going to use in this paper is the 2014 network, which features 2,464 nodes and 738,374 edges.  
  
\item[US Airports network]
The dataset contains information on the flights between a number of US airports during the year 2017. Each link represents a connection between airports, with the weight representing the number of passengers on all flights on that route in the given direction. The system contains 1151 airports and 20,580 different connections. The same network with data coming from different years has already been used in network filtering literature \cite{backbone_vespignani,gemmetto2017irreducible}. 
\end{description}

In Appendix \ref{sec:sn8} we show comparisons between the P\'olya filter and other filtering techniques on the two following additional datasets.
\begin{description}
\item[High School network]
This dataset reports face-to-face interactions between students recorded in 2013 in a Marseille high school throughout a period of five days \cite{mastrandrea2015contact}. The weights on the network's links correspond to the number of interactions recorded during the experiment, and interactions were recorded every 20 seconds. The network is made of 5818 weighted interactions among 1567 students.

\item[Florida ecosystem network]
Weights in this network represent the carbon exchanges between taxa in the cypress wetlands of South Florida during its dry season \cite{ulanowicz2005network}. The network is formed of 128 nodes and 2137 links.
\end{description}

\subsection{Approximations of the P\'olya filter's $p$-values and relationships with the disparity filter}

Eq. \eqref{eq:Polya_prob_main} can be considerably simplified assuming $s \gg k/a$, and $w \gg 1$. In this regime, the $p$-value the P\'olya filter associates to a weight $w$ on a link belonging to a node with degree $k$ and strength $s$ reduces to
\begin{equation} \label{eq:p_Polya_approx}
\pi_P (w \mid k,s,a) \approx \frac{1}{\Gamma\left[ \frac{1}{a} \right]} \left( 1- \frac{w}{s} \right)^{\frac{k-1}{a}} \left( \frac{w \, k}{s \, a} \right)^{\frac{1}{a}-1} \ ,
\end{equation}
where $\Gamma$ is the Gamma function. The rigorous derivation of the above approximation is provided in Appendix \ref{sec:sn3}, where we also show numerically that the approximations used to derive Eq. \eqref{eq:p_Polya_approx} hold for large fractions of edges. If we further approximate Eq. \eqref{eq:p_Polya_approx} by expanding it around $w / s \approx 0$ we obtain
\begin{equation}\label{eq:top_bound_p_Polya}
\pi_P \approx \frac{e^{-\frac{r}{a}} \left( \frac{r}{a} \right)^{\frac{1}{a}-1}}{\Gamma\left[ \frac{1}{a} \right]} \ ,
\end{equation}
where $r$ was introduced in Eq. \eqref{eq:r_def}. This result demonstrates the soft dependence of the P\'olya filter on the ratio $r$ mentioned in the main text and shown in Fig. \ref{fig:ratio_explained_1}.

Notably, when Eq. \eqref{eq:p_Polya_approx} holds, the P\'olya filter does not depend on $w$ and $s$ separately (as it normally does, as per Eqs. \eqref{eq:Polya_filter} and \eqref{eq:Polya_prob_main}), but only depends on such quantities through the ratio $w/s$ and the $p$-value loses its ability to discriminate between nodes with different heterogeneity. As we shall see in the following section, this allows to extend the applicability of the P\'olya filter to networks with non-integer weights.

\begin{figure}[!b]
\centering
\includegraphics[width=\linewidth]{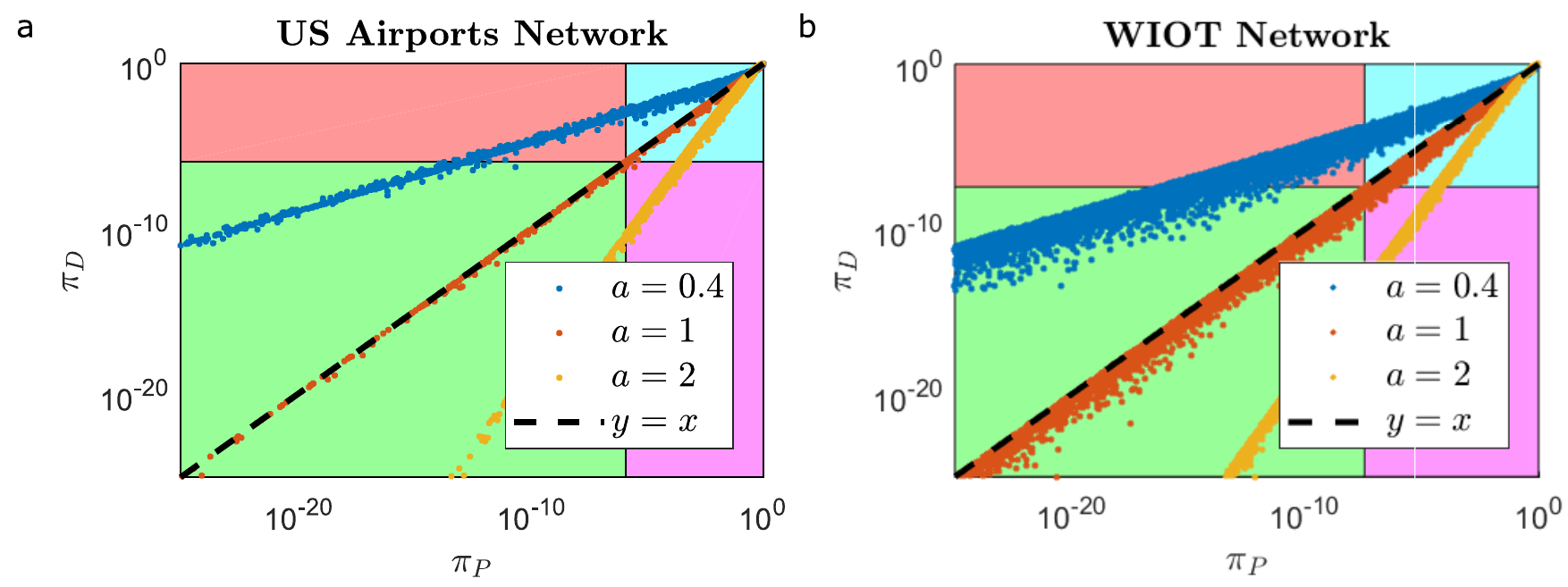}
\caption{Comparison of the $p$-values prescribed by the disparity ($\pi_D$) and P\'olya ($\pi_P$) filters computed for different values of $a$ (at a univariate significance level $\alpha_u = 0.05$). Each region of the plot is coloured depending on the significance of the two filters. Points in the blue (green) region correspond to links rejected (accepted) by both filters, while points in the purple (red) region correspond to links accepted only by the disparity (P\'olya) filter. (\textbf{a}) $p$-values computed on the US Airports network. (\textbf{b}) $p$-values computed on the US Airports network.}
\label{fig:Polya_vs_disparity}
\end{figure}

Setting $a=1$ in Eq. \eqref{eq:p_Polya_approx} gives $\pi_P = \left( 1- w / s \right)^{k-1}$, which coincides with the $p$-value prescribed by the disparity filter \cite{backbone_vespignani}, i.e.,
\begin{equation}\label{eq:disparity_cdf}
\pi_D (w | k,s) = 1- (k-1)\int_0^{w/s} (1-x)^{k-2} \,\,\mathrm{d}x \;=\; \left(1-\frac{w}{s}\right)^{k-1} \ .
\end{equation}
We can therefore conclude that the disparity filter corresponds to a large strength approximation of the P\'olya filter in a special case ($a=1$). This is demonstrated in Fig.~\ref{fig:Polya_vs_disparity}, where we plot the relationship between the $p$-values assigned by the P\'olya and disparity filters to the same links. As it can be seen, the two sets of values are indeed very close when $a=1$. This should not come as a surprise. Indeed, the null hypothesis underlying the disparity filter is ruled by a particular case of the Dirichlet distribution, which is known to be a limit case  of the Beta-Binomial distribution as the number of draws goes to infinity \cite{polya_dirichelet}. 

The relationship between the P\'olya and disparity filters is further investigated in Appendices \ref{sec:sn3} and \ref{sec:sn6}.

\subsection{Equivalence of P\'olya backbones}

In this Section we are going to show that the backbones produced by the P\'olya filter for different values of $a$ can be made approximately equivalent by tuning the filter's statistical significance.

Assessing the statistical significance of a link with weight $w$ (or associated to a value $r$ of the ratio in Eq. \eqref{eq:r_def}) entails determining whether it is compatible with the assumed null hypothesis. Using a Gaussian analogy, we can say that a value $r$ is compatible with the null hypothesis if $\mu_r(a) - b \sigma_r(k,s,a) < r < \mu_r(a) + b \sigma_r(k,s,a)$, where $b \geq 0$ is inversely proportional to the statistical significance $\alpha$, while $\mu_r$ and $\sigma_r$ denote the expected mean and standard deviation of the ratio $r$ under the P\'olya null hypothesis. These read:

\begin{alignat}{2}
\mu_r(a) &= \mathbb{E} \left[ r \right]= 1 \\
\sigma^2_r (k,s,a) &= \mathbb{E} \left[ (r-\mu_r )^2 \right]= \frac{k-1}{s} \,\, \frac{k+a s}{a+k} \ .
\end{alignat}

Let us then consider the null hypotheses associated with two different values $a_1$ and $a_2$ of the parameter, such that $a_2 \geq a_1$, and look for a scaling parameter $c$ that makes them equivalent. In order to do so we just need to impose:
\begin{equation}
\mu_r(a_1) \pm \sigma_r(k,s,a_1) = \mu_r(a_2) \pm c \sigma_r(k,s,a_2) \ ,
\end{equation}
for $c \geq 0$. Using $\mu_r(a_1) = \mu_r(a_2) = 1$, and setting $a_2 = d a_1$ (with $d \geq 1$), we can solve the above equation for $c$ and get
\begin{equation}
c = \sqrt{ \frac{a_1+k/d}{a_1+k}\frac{a_1 s+k}{a_1 s + k/d} } \; ,
\end{equation}
which is a monotonically decreasing function of $d$. This means that the same backbone produced by the P\'olya filter for $a=a_1$ can be approximately reproduced with $a = a_2 \geq a_1$ and a smaller region of compatibility with the null hypothesis (i.e., a higher statistical significance). In other words, in the P\'olya filter family of backbones, tolerance to heterogeneity and statistical significance are closely related. 

\subsection{Networks with non-integer weights}

The P\'olya filter is encoded in Eq.~\eqref{eq:Polya_filter}, which depends on $w$ and $s$ individually. This means, that Eq.~\eqref{eq:Polya_filter} is able to discriminate between nodes with different heterogeneity (given a fixed value of $k$), e.g., between two nodes characterised by the pairs $(w,s)=(10,100)$ and $(w,s)=(100,1000)$, respectively. This feature is naturally suited to deal with integer weights, such as those coming from counting experiments (e.g., as in the US Airports network).

The above property vanishes when $s \gg k/a$ and $w \gg 1$, leading to Eq.~\eqref{eq:p_Polya_approx}, which only depends on the ratio $w/s$ and, in fact, should be exploited to apply the P\'olya filter when dealing with networks with non-integer weights, even in cases when such approximations do not hold. Of course, doing so will change the underlying null hypothesis: indeed, Eq.~\eqref{eq:p_Polya_approx} does not assign a $p$-value to a weight $w$, but rather to a rate of interaction $w/s$. In most cases the $p$-values given by Eq.~\eqref{eq:Polya_filter} and Eq.~\eqref{eq:p_Polya_approx} are practically the same (see Fig. \ref{fig: polya approximation}), and can be used interchangeably when dealing with integer weights. Conversely,  Eq.~\eqref{eq:Polya_filter} cannot assign $p$-values to non-integer weights, but in such cases one can always assign a $p$-value to the interaction rate $w/s$ through Eq.~\eqref{eq:p_Polya_approx}.

We can further justify the use of Eq.~\eqref{eq:p_Polya_approx} by thinking of an overall rescaling of the weights by a large factor $c$. For example, let us consider a network whose lowest weights are of order $10^{-4}$. Applying Eq.~\eqref{eq:Polya_filter} to such a network would entail rescaling its weights by a factor $c \geq 10^4$ before filtering. Doing so, however, automatically takes us to the regime under which Eq.~\eqref{eq:p_Polya_approx} holds (i.e., $s \gg k/a$ and $w \gg 1$), which therefore becomes the P\'olya filter's analytical expression for non-integer weights.

\section{Explicit expression for the P\'olya filter's $p$-value} \label{sec:sn1}

The sum in Eq. \eqref{eq:Polya_prob_main} can be computed explicitly in order to derive an explicit expression for the $p$-value assigned by the P\'olya filter to a link with weight $w$ attached to a node with strength $s$ and degree $k$. This reads: 
\begin{equation} 
\label{eq:Polya_prob}
\begin{aligned}
\pi_P (w \mid k,s,a) &= && 1- \sum_{x=0}^{w-1} \mathbb{P}(x \mid k,s,a ) = \\
&= &&  \frac{B\left(\frac{k-1}{a}+s-w,w+\frac{1}{a}\right)}{(s+1) B\left(\frac{1}{a},\frac{k-1}{a}\right) B(s-w+1,w+1)} \times\\
&  &&\times \, _3F_2\left[ \begin{aligned} 
& 1,w+\frac{1}{a},-s+w \\
& w+1,-\frac{k-1}{a}-s+w
\end{aligned} ;1 \right] \;,
\end{aligned} 
\end{equation}
where $B$ is the Beta function, and $_3F_2$ denotes the generalised hypergeometric function.

It should be noted, however, that the above expression is of little practical use from the numerical viewpoint, due to the presence of the generalised hypergeometric function. Indeed, computing the $p$-values of the P\'olya filter through the sum of the probabilities reported in Eq. \eqref{eq:Polya_prob} is both faster and more accurate, as values of the beta function can be easily computed by any numerical software with high accuracy. Yet, the above expression is useful to gain analytical insight into the P\'olya filter. As a matter of fact, we shall use it in Appendix \ref{sec:sn3} to derive useful approximations and to prove the relationship between the P\'olya and disparity filters.

\section{The P\'olya filter for directed weighted networks} \label{sec:sn2}

Systems where the directionality of interactions cannot be neglected are usually described in terms of directed weighted networks \cite{totale_barabasi,totale_Newman}. The difference between weighted directed and weighted undirected networks is that the former are described in terms of a symmetric adjacency matrix $W$ such that $w_{ij}=w_{ji}, \ \forall \, i,j$, where the activity of each node can be specified in terms of a single degree $k_i = \sum_{j} \mathbb{1} (w_{ij})$ or strength $s_i = \sum_{j} w_{ij}$. The latter are instead formalized in terms of non-symmetric adjacency matrices, which requires to specify the in- and out-degrees ($k_i^\mathrm{in} = \sum_{j}\mathbb{1}(w_{ji})$ and $k_i^\mathrm{out} = \sum_{j} \mathbb{1} (w_{ji})$, respectively), and the in- and out-strengths ($s_i^\mathrm{in} = \sum_{j} w_{ji}$ and $s_i^\mathrm{out} = \sum_{j} w_{ji}$, respectively) for each node.  

The P\'olya filter can be easily generalised to weighted directed networks. In the undirected case each weight can be associated with two $p$-values, one for each of the two nodes the link is attached to. In the directed case we can still associate two $p$-values to each weight by assessing its statistical significance both as an incoming and as an outgoing link. For example, when testing as an outgoing link, Eq. \eqref{eq:Polya_prob} is easily generalized as (we drop all node indices to keep notation light)
\begin{eqnarray} \label{eq: polya prob out}
\pi_P (w \mid k^\mathrm{out}, s^\mathrm{out}, a) =  \frac{B\left(\frac{k^\mathrm{out}-1}{a}+s^\mathrm{out}-w,w+\frac{1}{a}\right)}{(s^\mathrm{out}+1) B\left(\frac{1}{a},\frac{k^\mathrm{out}-1}{a}\right) B(s^\mathrm{out}-w+1,w+1)} \times \\ \times \, _3F_2\left[ \begin{aligned} 
& 1,w+\frac{1}{a},-s^\mathrm{out}+w \\
& w+1,-\frac{k^\mathrm{out}-1}{a}-s^\mathrm{out}+w+1 
\end{aligned} ;1 \right] \ ,
\end{eqnarray}
with the replacements $k^\mathrm{out} \rightarrow k^\mathrm{in}$, $s^\mathrm{out} \rightarrow s^\mathrm{in}$ for the test as an incoming link. Both $p$-values can be tested against the same univariate threshold $\alpha$. A link is retained by the P\'olya filter only when at least one of the two $p$-values is lower than $\alpha$.

A link is kept only if at least one of the two $p$-values is lower than $\alpha_B$. In the case where $k_i^\mathrm{out}=1$, we keep the directed link connecting $i$ and $j$ only if $\pi_P (w_{ij} \mid k_j^\mathrm{in}, s_j^\mathrm{in}, a) < \alpha_B$, and vice versa in the case $k_j^\mathrm{in}=1$.

\section{Generalizing the disparity filter} \label{sec:sn3}

In this section we explicitly show how the disparity filter \cite{backbone_vespignani} can be recovered as a special case of the P\'olya filter for $a=1$. We start by rewriting the $p$-value associated with a weight $w$ attached to a node with degree $k$ and strength $s$. For the sake of simplicity, we go back to the undirected case of Eq. \eqref{eq:Polya_prob}:

\begin{equation} \label{eq: polya prob}
\pi_P (w \mid k,s,a) = \frac{B\left(\frac{k-1}{a}+s-w,w+\frac{1}{a}\right)}{(s+1) B\left(\frac{1}{a},\frac{k-1}{a}\right) B(s-w+1,w+1)} 
\, _3F_2\left[ \begin{aligned} 
& 1,w+\frac{1}{a},-s+w \\
& w+1,-\frac{k-1}{a}-s+w
\end{aligned} ;1 \right] \ .
\end{equation}

In the following, we will repeatedly simplify the above expression by making use of the zero-order Stirling approximation for the ratio of two Gamma functions:
\begin{equation}\label{eq: gamma expansion}
\frac{\Gamma \left[ x + \alpha \right] }{\Gamma \left[ x + \beta \right]} = x^{\alpha - \beta} \left( 1 + \mathcal{O}\left[ \frac{1}{x} \right] \right) \approx x^{\alpha - \beta} \ ,
\end{equation}
which holds for $x \to \infty$. 

\begin{figure}[!htbp]
	\centering
	\includegraphics[width=0.45\linewidth]{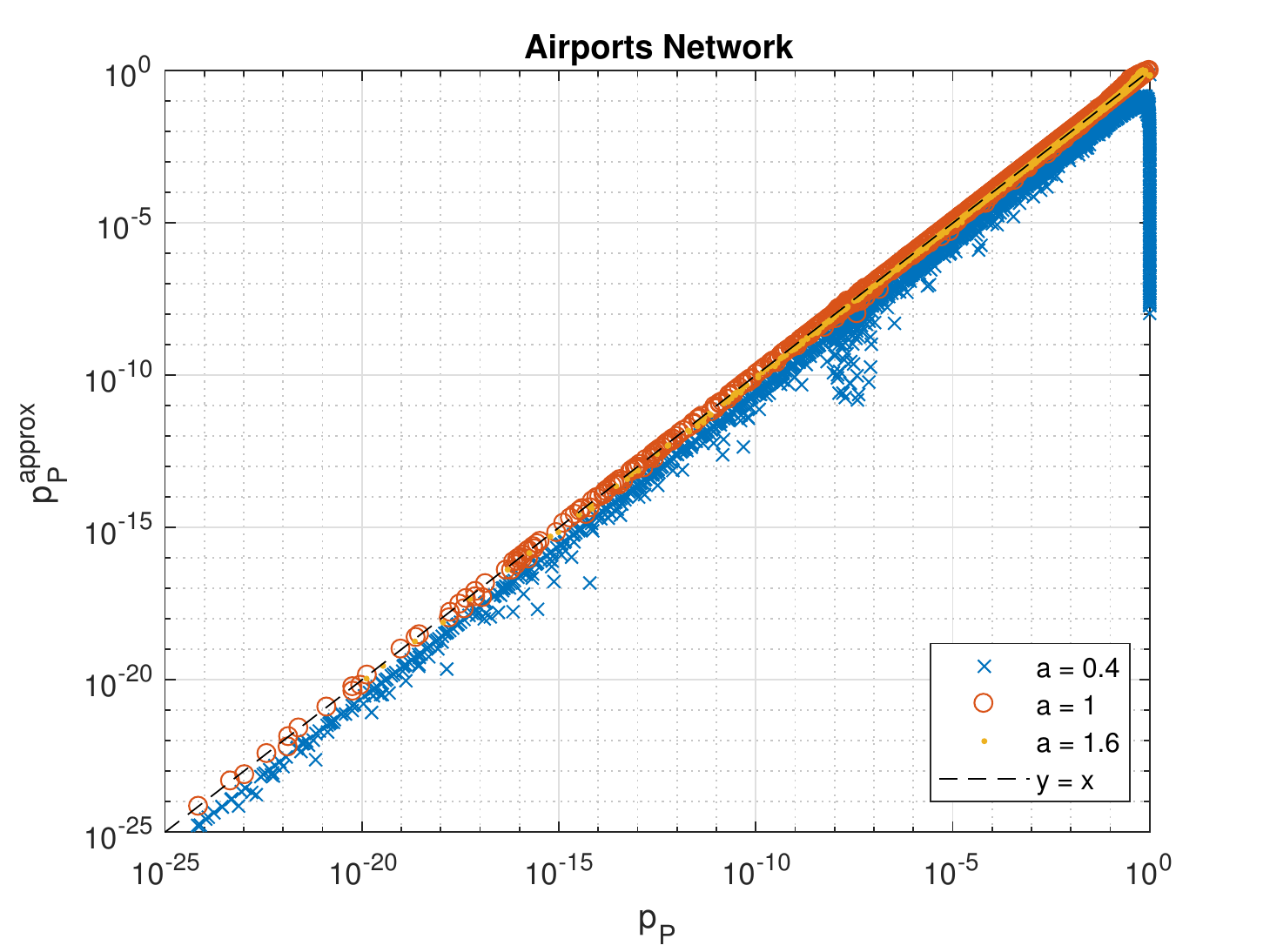}	
	\includegraphics[width=0.45\linewidth]{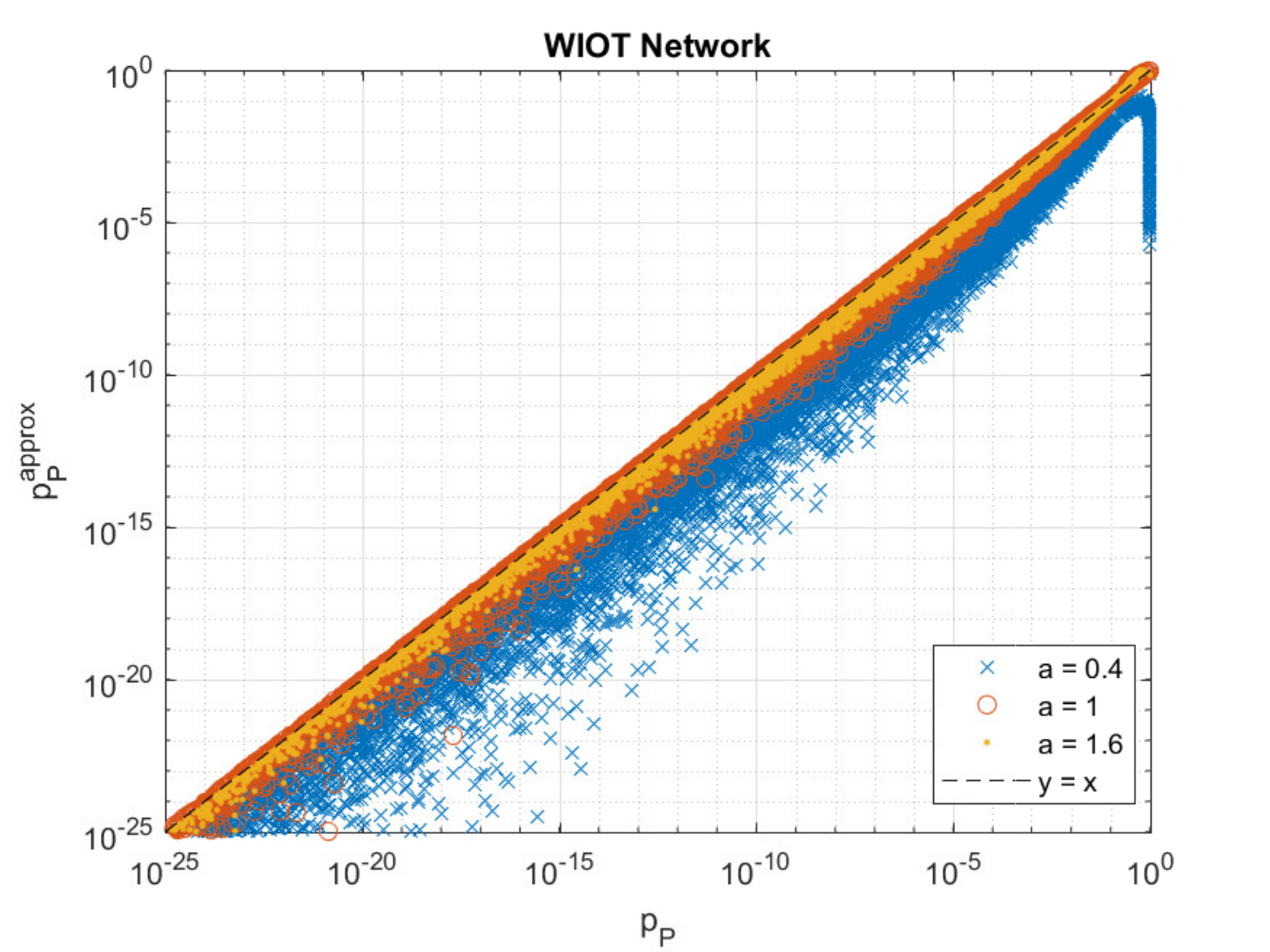}	
	\caption{Scatter plots of the $p$-values obtained from the P\'olya filter compared with the approximate expression in Eq. \eqref{eq: p polya approx} for different values of the parameter $a$.
	}
	\label{fig: polya approximation}
\end{figure}

We first take care of the hypergeometric function in Eq. \eqref{eq: polya prob}. We start by expanding it in terms of ratios of Gamma functions:
\begin{eqnarray}\label{eq: hypergeom simplified 1 }
&& _3F_2\left[ \begin{aligned} 
& 1,w+\frac{1}{a},-s_i+w \\
& w+1,-\frac{k-1}{a}-s+w 
\end{aligned} ;1 \right] = \\ &=& \sum_{n=0}^{\infty}  \frac{\Gamma\left[ -s+w+n \right]}{\Gamma\left[ -s+w \right]} \; \frac{\Gamma\left[ -\frac{k-1}{a}-s+w+1 \right]}{\Gamma\left[ -\frac{k-1}{a}-s+w+1+n \right]} \frac{\Gamma\left[ w+\frac{1}{a}+n \right]}{\Gamma\left[w+\frac{1}{a}\right]} \frac{\Gamma\left[ w+1 \right]}{\Gamma\left[w+1+n\right]} \ .
\end{eqnarray}
We can simplify the last two terms in the above expression:
\begin{equation*}
\frac{\Gamma\left[ w+\frac{1}{a}+n \right]}{\Gamma\left[w+1+n\right]} \frac{\Gamma\left[ w+1 \right]}{\Gamma\left[w+\frac{1}{a}\right]} \approx w^{\frac{1}{a}+n-(1+n)} w^{1-\frac{1}{a}} = 1 \ ,
\end{equation*}
where we have assumed $w \gg 1 /a$. Putting this result back into Eq. \eqref{eq: hypergeom simplified 1 } gives:
\begin{equation}\label{eq: hypergeom simplified 2 }
_3F_2\left[ \begin{aligned} 
& 1,w+1+\frac{1}{a},-s+w+1 \\
& w+2,-\frac{k-1}{a}-s+w+2 
\end{aligned} ;1 \right] \approx \; _2F_1\left[ \begin{aligned} 
& -s+w,1 \\
& -\frac{k-1}{a}-s+w+1 
\end{aligned} ;1 \right] \ .
\end{equation}
Eq. \eqref{eq: hypergeom simplified 2 } can be now further simplified by making use of the the Chu-Vandermonde identity $_2F_1(-n,b;c,1) = \frac{(c-b)_n}{(c)_n}$ (where $(\cdot)_n$ denotes the Pochhammer symbol), which gives:
\begin{equation}\label{eq: hypergeom simplified 3 }
_2F_1\left[ \begin{aligned} 
& -s+w,1 \\
& -\frac{k-1}{a}-s+w+1 
\end{aligned} ;1 \right] \, = \, \frac{s-w+\frac{k-1}{a}}{(k-1)/a} \ .
\end{equation}
Putting Eq. \eqref{eq: hypergeom simplified 3 } back into Eq. \eqref{eq: polya prob}, and writing the Beta functions in Eq. \eqref{eq: polya prob} as ratios of Gamma functions, allows to write Eq. \eqref{eq: polya prob} as the product of the three following ingredients:
\begin{equation}
\begin{aligned}
& B\left[\frac{k-1}{a}+s-w,w+\frac{1}{a}\right](s-w+\frac{k-1}{a}) &&= \frac{\Gamma\left[ \frac{k-1}{a}+s-w+1 \right] \Gamma\left[ w+\frac{1}{a} \right]}{\Gamma\left[ s+\frac{k}{a}\right]} \\
& \frac{1}{(s+1) B\left[s-w+1,w+1\right]} &&= \frac{\Gamma\left[ s+1 \right]}{\Gamma\left[ s-w+1\right]\Gamma\left[ w+1\right]} \\
& \frac{1}{\frac{k-1}{a} B\left[\frac{1}{a},\frac{k-1}{a}\right]} &&= \frac{\Gamma\left[ \frac{k}{a} \right]}{\Gamma\left[ \frac{1}{a}\right]\Gamma\left[ \frac{k}{a}- \frac{1}{a} +1 \right]} \ .
\end{aligned}
\end{equation}
By matching Gamma functions in the numerators and denominators of the above ratios, and making use of the Stirling approximation (Eq. \eqref{eq: gamma expansion}), we can then write down the $p$-value in Eq. \eqref{eq: polya prob} as the product of the following quantities:
\begin{equation}\label{eq: all gamma approx}
\begin{aligned}
& \frac{\Gamma\left[ s -w + \frac{k-1}{a}+1 \right]}{\Gamma\left[ s-w+1 \right]} && \approx \;\; \left( s-w \right)^{\frac{k-1}{a}} = s^{\frac{k-1}{a}} \left( 1- \frac{w}{s} \right)^{\frac{k-1}{a}} \; , \qquad s-w \gg \frac{k-1}{a}+1 \\
& \frac{\Gamma\left[w+ \frac{1}{a} \right]}{\Gamma\left[ w+1 \right]} && \approx \;\; w^{\frac{1}{a}-1} \; , \qquad w \gg \frac{1}{a} \, , \, w \gg 1 \\
& \frac{\Gamma\left[s+1 \right]}{\Gamma\left[ s+\frac{k}{a} \right]} && \approx \;\; s^{1-\frac{k}{a}} \; , \qquad s \gg \frac{k}{a} \, , \, s \gg 1 \\
& \frac{\Gamma\left[\frac{k}{a} \right]}{\Gamma\left[ \frac{k}{a}-\frac{1}{a}+1 \right]} && \approx \;\; \left( \frac{k}{a} \right)^{\frac{1}{a}-1} \; , \qquad k \gg a-1 \ ,
\end{aligned}
\end{equation}
where on each line we have written the approximations we made use of. Finally, we can put together the above expressions, which gives the result reported in Eq. \eqref{eq:p_Polya_approx}:
\begin{equation}\label{eq: p polya approx}
\pi_P (w \mid k,s,a) \approx \frac{1}{\Gamma\left[ \frac{1}{a} \right]} \left( 1- \frac{w}{s} \right)^{\frac{k-1}{a}} \left( \frac{w \, k}{s \, a} \right)^{\frac{1}{a}-1} \ .
\end{equation}
All the approximations that we are assuming are written in Eq. \eqref{eq: all gamma approx}. In Fig. \ref{fig: polya approximation} we show a comparison between the $p$-values obtained from the P\'olya filter (Eq. \eqref{eq: polya prob}) and the above expression in the two networks we consider. As it can be seen, the overall agreement is rather good, and larger values of $a$ improve the quality of the approximation, as it can be seen from Eq. \eqref{eq: all gamma approx}.

\section{Thresholding on $r$} \label{sec:sn4}

As discussed above (see  Eq. \eqref{eq:top_bound_p_Polya}), there is a soft relationship between the value of the $r$ ratio of a link and the corresponding $p$-value assigned by the P\'olya filter to it. In short, links associated with high values of $r$ tend to be retained, but the opposite does not necessarily hold, i.e., links associated to low values of $r$ can still be validated by the filter and contribute to the overall heterogeneity of P\'olya backbones.

\begin{figure}[h!]
\centering
\includegraphics[width=.45\linewidth]{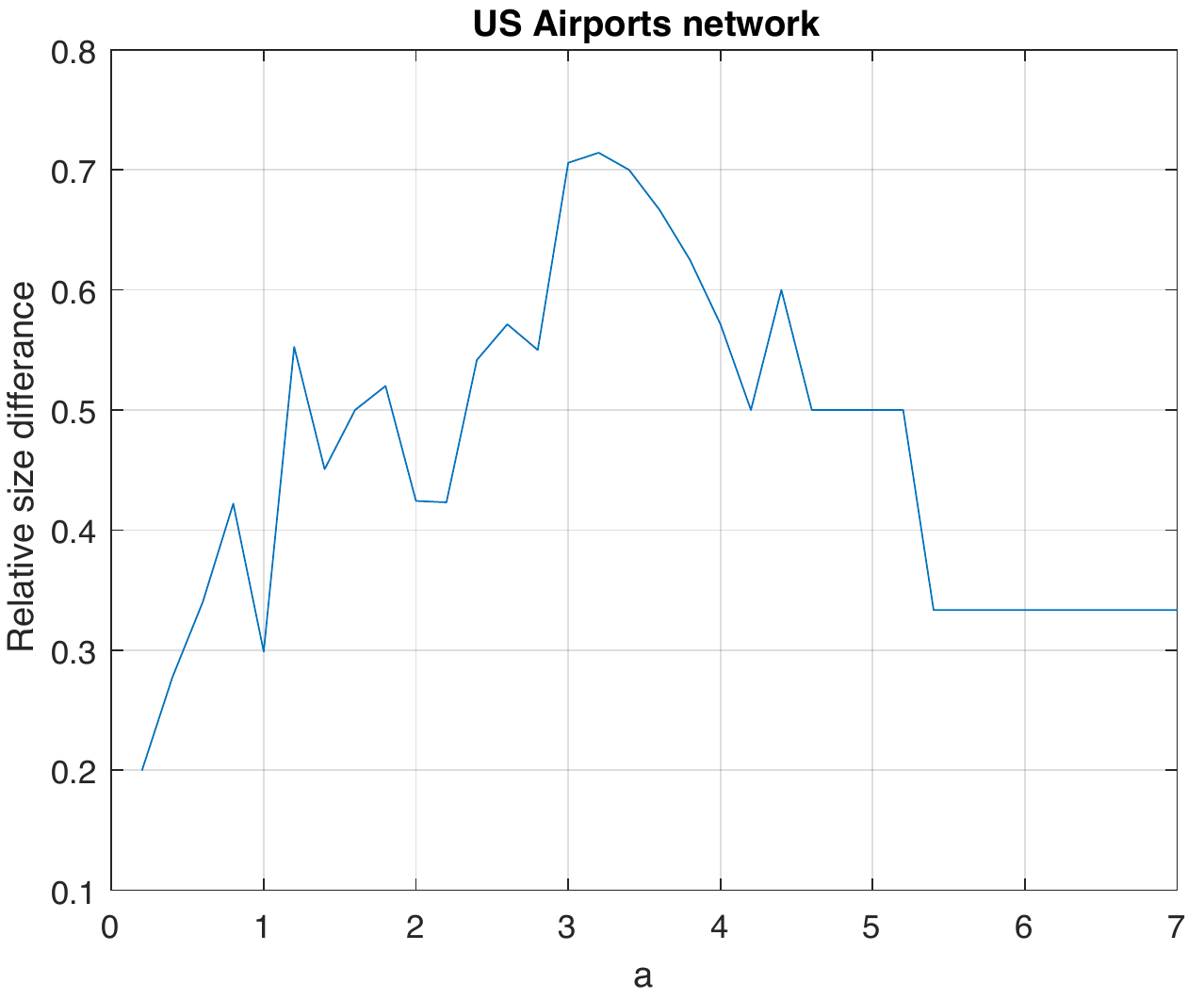}
\includegraphics[width=.45\linewidth]{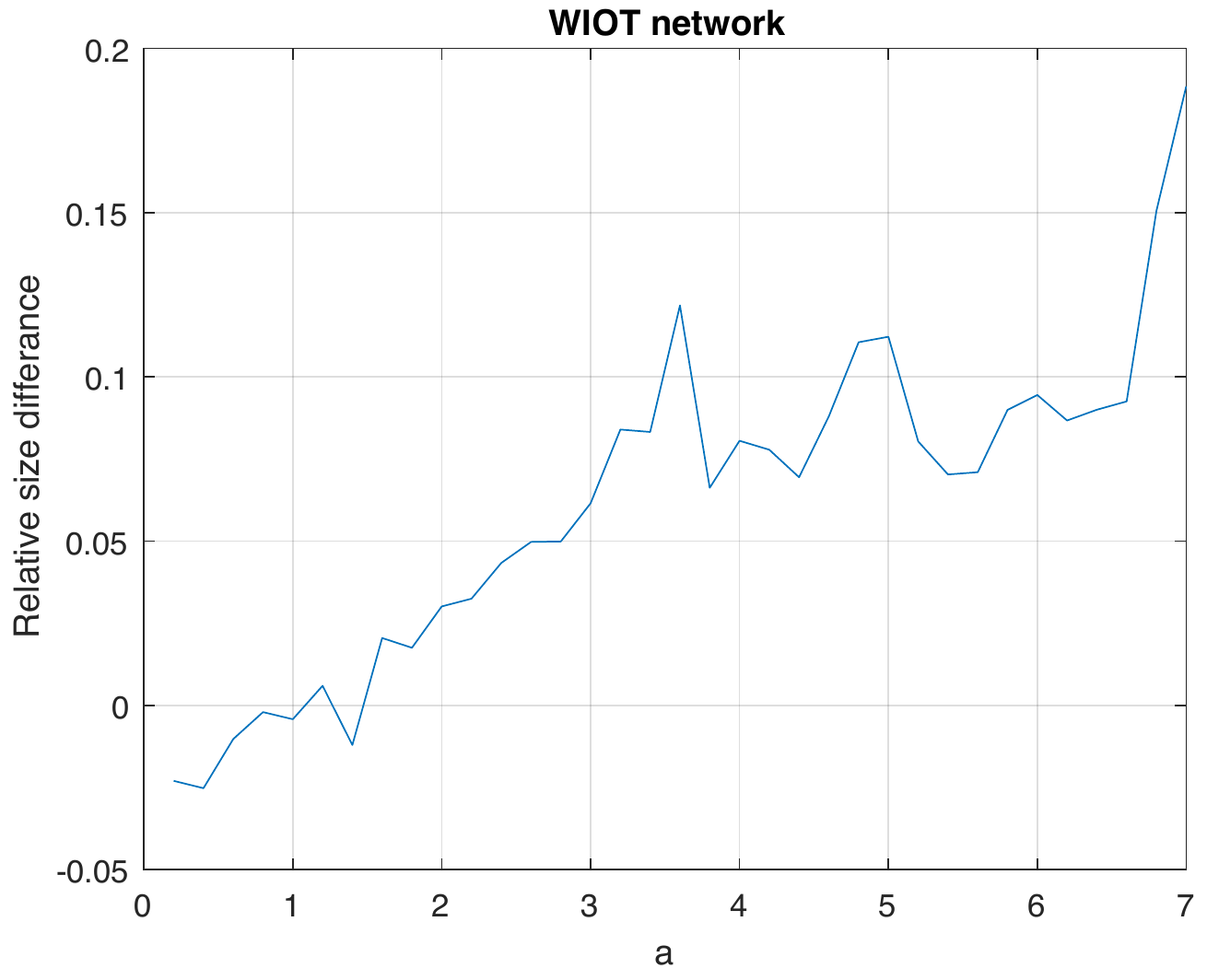}
\caption{Relative difference in the size of the network's largest connected component as measured in the full P\'olya backbone and in the backbone obtained by thresholding on $r$ via Eq. \eqref{eq:r_thr}.}
\label{fig:thr_r}
\end{figure}

In order to highlight this point, in Fig. \ref{fig:thr_r} we plot the relative difference between the largest connected components of full P\'olya backbones, and those of the backbones that would be obtained by thresholding on $r$. Thresholding is performed by inverting Eq. \eqref{eq:top_bound_p_Polya} in order to determine the value $r_\mathrm{thr}$ such that 
\begin{equation} \label{eq:r_thr}
\alpha_B = \frac{e^{-\frac{r_\mathrm{thr}}{a}} \left ( \frac{r_\mathrm{thr}}{a} \right )^{\frac{1}{a}-1}}{ \Gamma \left [ \frac{1}{a} \right ]} \ ,
\end{equation}
where $\alpha_B$ is the Bonferroni-corrected multivariate significance level adopted to filter. As it can be seen, both in the case of the US air transport and WIOT networks, thresholding leads to backbones that are considerably more disconnected. This is somewhat to be expected, since thresholding implies producing sparser backbones by discarding links with $r < r_\mathrm{thr}$ that might be instead validated by the full P\'olya filter. Yet, as is particularly apparent in the US air transport network, the sparsification of the largest connected component can be very significant.

The main reason behind this lies in the fact that links associated with high values of $r$ are typically those with a large weight $w$ or those attached to a hub (i.e., with a high $k$). As such, these links can be easily expected to be validated, unless the parameter $a$ is increased to the point where the network's own heterogeneity is used as null hypothesis (see, for example, the case study on US air transport network, where all links connecting major hubs are filtered out when setting $a = a_\mathrm{ML}$). Conversely, links with lower values of $r$ that are still validated by the P\'olya filter correspond to statistically significant \emph{combinations} of $w$, $k$, and $s$, which contribute to the heterogeneity of P\'olya backbones (see Appendix \ref{sec:sn6}).

\section{Maximum Likelihood Estimates}\label{sec:sn5}

As a parametric approach, the P\'olya filter lends itself to optimization procedures aimed at identifying the value of the parameter $a$ most suited to the particular network under study. As mentioned previously, maximium-likelihood estimation (MLE) is a natural option to single out the ``nullest'' model in the P\'olya family for the network under consideration.

This can be achieved by solving
\begin{equation} \label{eq: like prob}
a_\mathrm{ML} = \mathrm{arg}\max_{a \in [0,\infty)} \mathcal{L} (a; \boldsymbol{w}) \ ,
\end{equation}
where $\boldsymbol{w}$ denotes the sequence of weights in the network, and
\begin{equation} \label{eq: log like func beta bin}
\mathcal{L} (a ; \boldsymbol{w}) = \sum_{i,j = 1}^N \log \mathbb{P}( w_{i j} \mid s_{i},k_i ) = \sum_{i,j = 1}^N \log \left[ \binom{s_i}{w_{i j}} \frac{B(\frac{1}{a} + w_{i j}, \frac{k_i-1}{a} + s_i- w_{i j})}{B(\frac{1}{a}, \frac{k_i-1}{a})} \right]
\end{equation}
is the log-likelihood function associated with the probability of observing the particular weight sequence under a P\'olya process with parameter $a$.

Solving the optimization problem in \eqref{eq: like prob} with the above function boils down to solving numerically the following equation:
\begin{equation} \label{eq: log like eq beta bin}
\begin{split}
\sum_{i,j=1}^N \left[-(k_i-1) \psi \left(\frac{k_i+a s_i-a w_{i j}-1}{a}\right)+k_i \psi \left(\frac{k_i}{a}+s_i\right)+  \right. \\ \left.
(k_i-1) \psi \left(\frac{k_i-1}{a}\right)-k_i \psi \left(\frac{k_i}{a}\right)-\psi \left(w_{i j}+\frac{1}{a}\right)+\psi \left(\frac{1}{a}\right) \right] = 0 \ ,
\end{split}
\end{equation}
where $ \psi(x) = \Gamma ' (x) / \Gamma(x)$ is the Polygamma function of order $0$. 

\begin{figure}
\centering
\includegraphics[width=.45\linewidth]{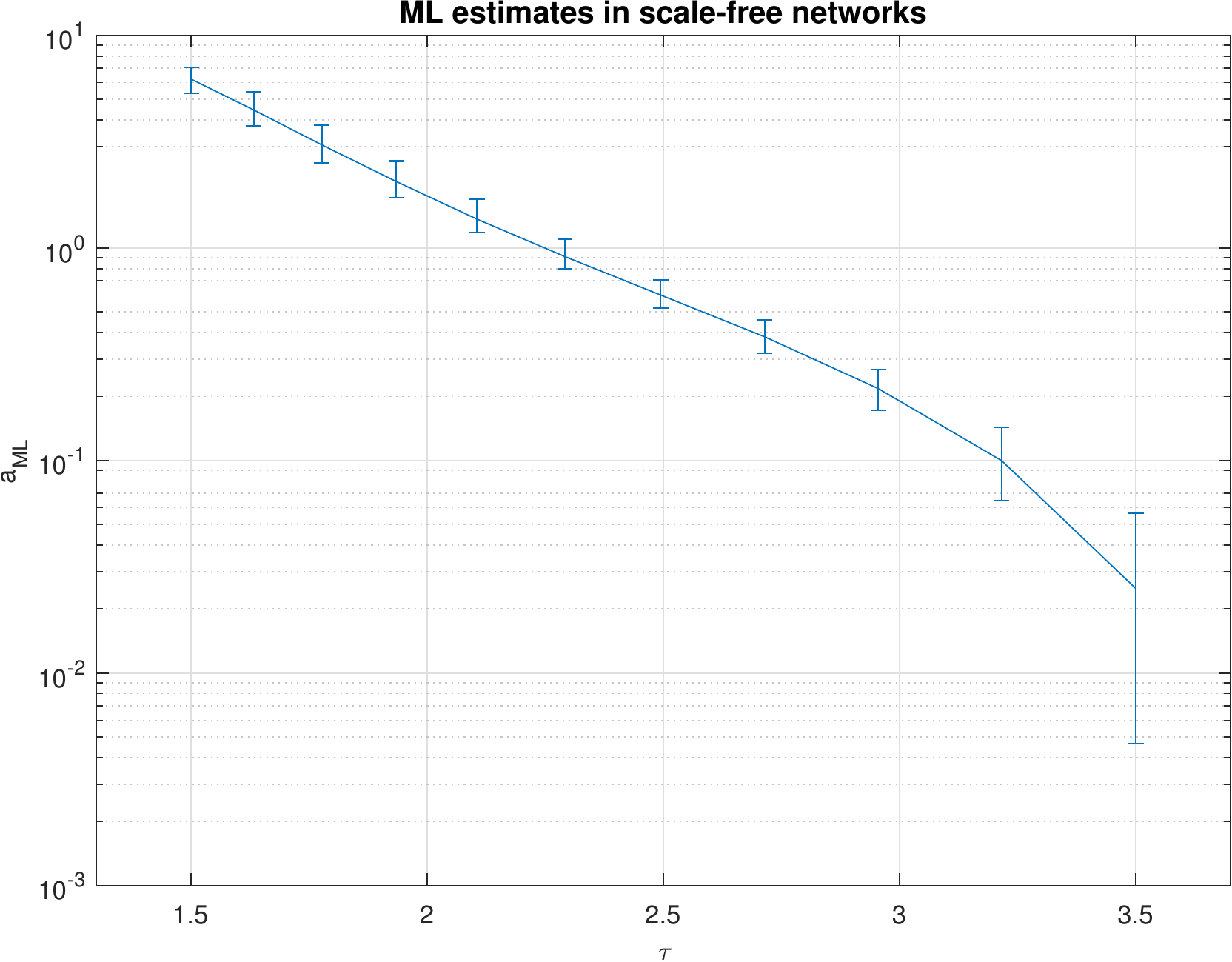}
\includegraphics[width=.45\linewidth]{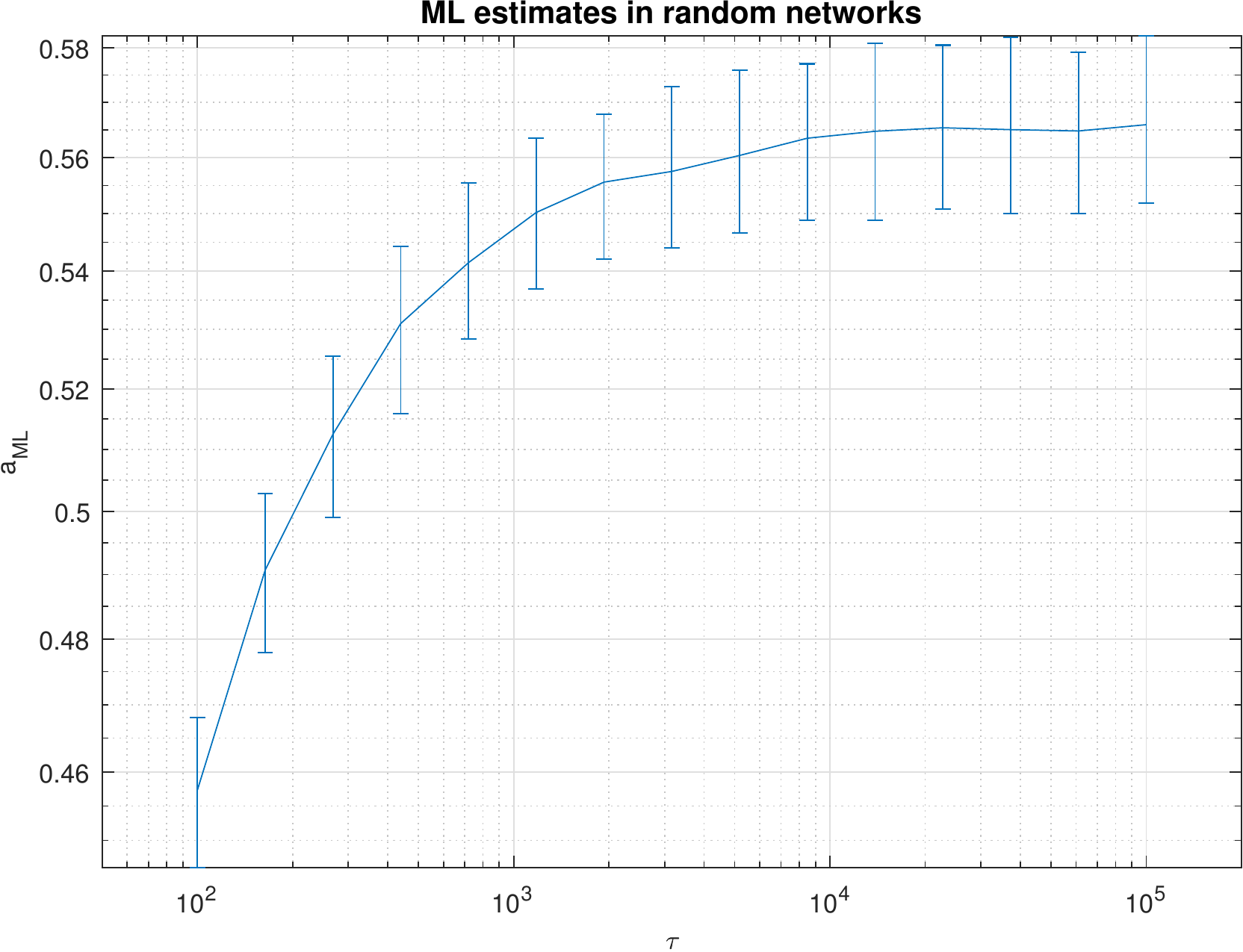}
\caption{ML estimates of the P\'olya filter's parameter $a$. In both cases, the networks are made of $3,000$ nodes and have an average degree of $8$. The error bar are $95\%$ confidence intervals obtained through $200$ different randomizations of both weights and topology. \textbf{Left panel}: ML estimates $a^*$ for Barabasi-Albert networks with a power-law weight distribution with tail exponent $\tau$. \textbf{Right panel}: ML estimates $a^*$ for Erd\H os-R\'enyi networks with uniform distribution of weights $U \left[ 1, \tau \right]$.}
\label{fig:MLestimations}
\end{figure}

In Fig. \ref{fig:MLestimations} we report ML estimates obtained on synthetic networks. The networks employed in the left panel are characterised by a scale-free topology generated using the BA model \cite{totale_barabasi} and a power-law weight distribution with tail exponent $\tau$. The optimal values $a_\mathrm{ML}$ clearly show that ML estimates respond to the network's heterogeneity, spanning almost three orders of magnitude ranging from values $a_\mathrm{ML} \simeq 10$ in the presence of very strong heterogeneity ($\tau = 1.5$) to $a_\mathrm{ML} \simeq 10^{-3} \text{--} 10^{-2}$ in the presence of mild heterogeneity. In the right panel of Fig. \ref{fig:MLestimations} we report the ML estimates on Erd\H os-R\'enyi random graphs with a uniform weight distribution $U \left[ 1, \tau \right]$, with weights rounded to the nearest integer. As it can been, the estimates are much less sensitive to changes with respect to the previous case, with $a_\mathrm{ML} \simeq 0.46 \text{--} 0.56$, which implies the \emph{de facto} impossibility to discriminate even between substantially different models when no marked heterogeneity is present in their weight distributions. 

\section{Comparing different P\'olya backbones} \label{sec:sn6}

\begin{figure}[h!]
\centering
\includegraphics[width=.45\linewidth]{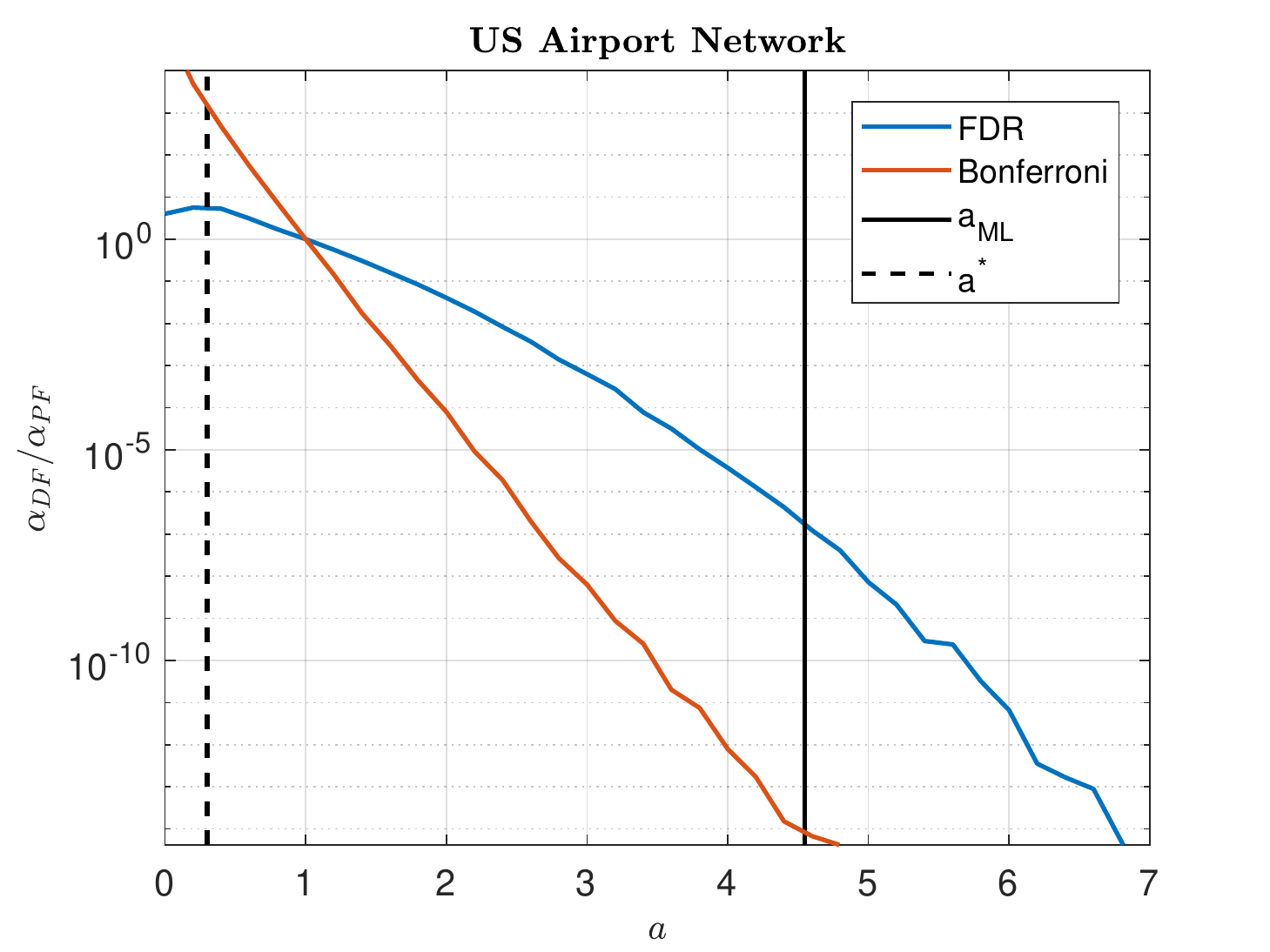}
\includegraphics[width=.45\linewidth]{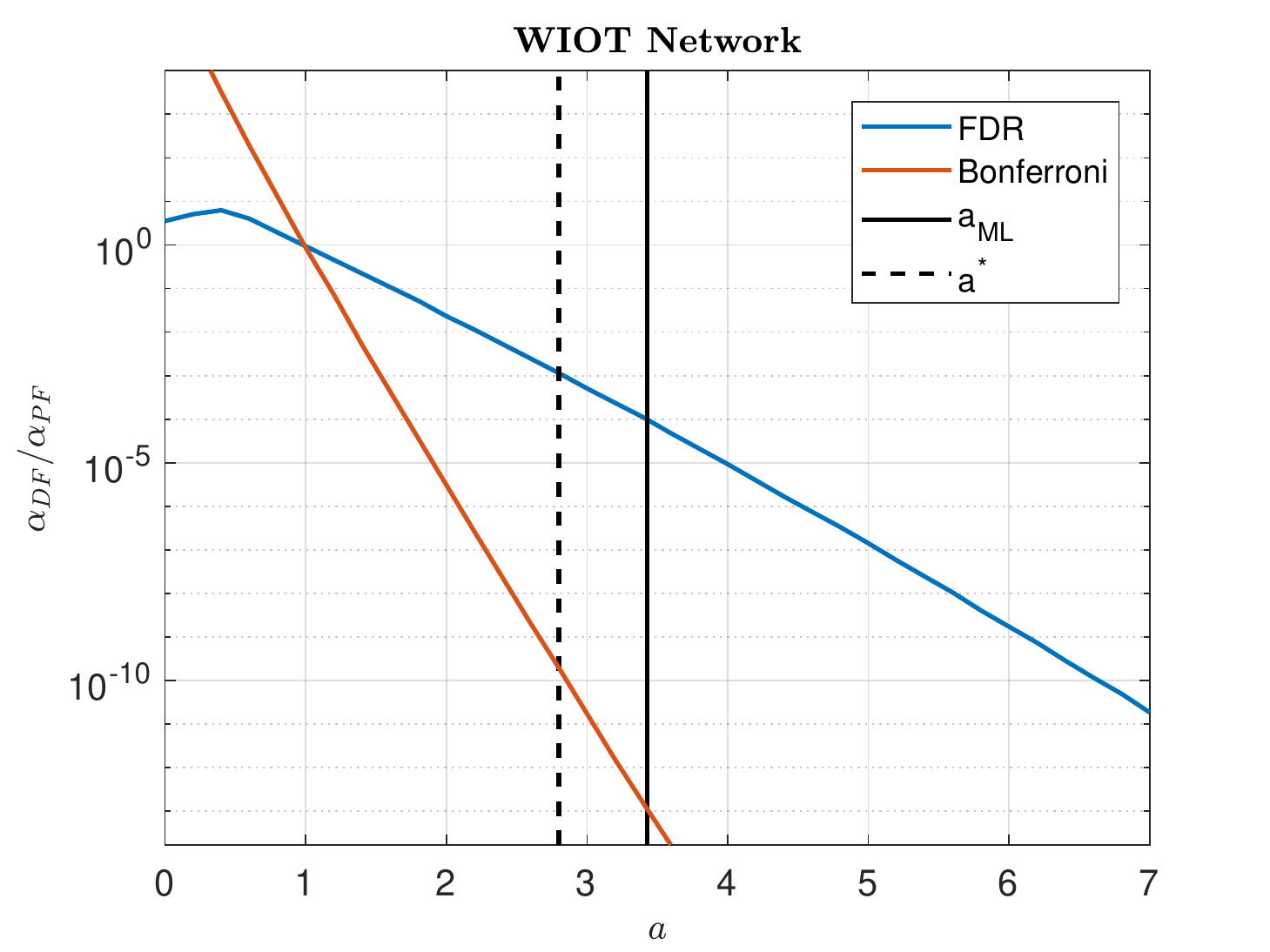}
\caption{Univariate statistical significance level $\alpha_\mathrm{DF}$ that has to be set for a disparity filter in order to match the backbones generated by P\'olya filters with different values of $a$ at a univariate significance level $\alpha_\mathrm{PF} = 0.05$. The lines correspond to the ratio $\alpha_\mathrm{DF} / \alpha_\mathrm{PF}$ when applying the Bonferroni (orange) and false discovery rate (blue) multiple test corrections. The solid vertical line corresponds to the maximum-likelihood value $a_\mathrm{ML}$ (see Appendix \ref{sec:sn5}), while the dashed vertical line corresponds to the one maximising the salience-related measure $O_1$ defined in Eq. \eqref{eq:optimality}.}
\label{fig:a_vs_alpha}
\end{figure}

In this Section we provide numerical evidence in support of the discussion in Appendix \ref{sec:methods} (``Equivalence of different P\'olya backbones''), where we argued that the backbones produced by the P\'olya filter at different values of $a$ can be made approximately equivalent by tuning the filter's statistical significance. \ref{fig:a_vs_alpha} shows the univariate statistical significance level $\alpha_\mathrm{DF}$ that has to be set for a P\'olya filter with $a=1$ (which closely approximates the disparity filter, as demonstrated in the main text) to match the backbones generated by P\'olya filters with different values of $a$ at a univariate significance level $\alpha_\mathrm{PF} = 0.05$.  

As it can be seen, regardless of the multiple testing correction applied (i.e., Bonferroni or FDR), the univariate thresholds required to make the backbones equivalent can differ by several orders of magnitudes. This is true, in particular, in correspondence of notable value of $a$, i.e., for $a=a^*$ (which denotes the value such that the salience-related metrics $O_1$ defined in Eq. \eqref{eq:optimality} is maximised) and for $a=a_\mathrm{ML}$ (i.e., when the network's own heterogeneity is used as benchmark for the P\'olya null hypothesis). 

All in all, these results show that P\'olya filters corresponding to different values of $a$ can be made equivalent by tuning their statistical significance. Yet, the above plots show that a difference in $a$ of a few units can lead to dramatic differences in terms of statistical significance (i.e., of ten or orders of magnitude or more). This, in turn, means that the same set of links can have drastically different statistical meanings when generated by different P\'olya filters. Indeed, decreasing the univariate threshold $\alpha$ by several orders of magnitude lowers the filter's tolerance to false positives by the same amount, while also causing a much higher false negative. Therefore, a link discarded by the P\'olya filter with parameter $a_2$ can can still be discarded by the P\'olya filter with parameter $a_1 < a_2$ (i.e., a lower tolerance to heterogeneity), but only by making the test extremely conservative.

\section{Relationship with salience} \label{sec:sn7}

Link salience is a recently introduced measure of link importance \cite{salience}, based on the distance between nodes. Given the adjacency matrix $W$ of weighted directed network, where an element $w_{i j}$ represent the strength of the interaction between nodes $i$ and $j$, the salience is computed through the auxiliary distance matrix $D$ such that $d_{i j} = 1/w_{i j}$ if $w_{i j}>0$ and $0$ otherwise. Once $D$ is known, the salience of a connection $(i,j)$ can be obtained. For a fixed reference node $r$, the set of weighted shortest paths to all other nodes is called the shortest-path tree matrix $T(r)$, which collects the most effective routes from $r$ to the rest of the network. $T(r)$ is a symmetric $N \times N$ matrix $T(r)$ such that $t_{i j}(r)=1$ if the link $(i,j)$ is part of \emph{at least} one of the shortest paths starting from $r$ and $t_{i j}(r)=0$ otherwise. Once all the possible $T(r) \;\; r=1,2 \ldots N$ matrices have been calculated, the salience of a link $(i,j)$ can be computed as:
\begin{equation}
S_{i j} = \frac{1}{N} \sum_{r=1}^{N} t_{i j}(r) \ .
\end{equation}
For a large collection of complex networks, it has been found that the distribution of link salience exhibits a peculiar bimodal shape in the unit interval, with most links ending up with $S \approx 0$ or $S \approx 1$. As a result, salience could be used to extract a network backbone, as this would practically not be affected by any particular salience threshold. 

\begin{figure}
\centering
\includegraphics[width=.45\linewidth]{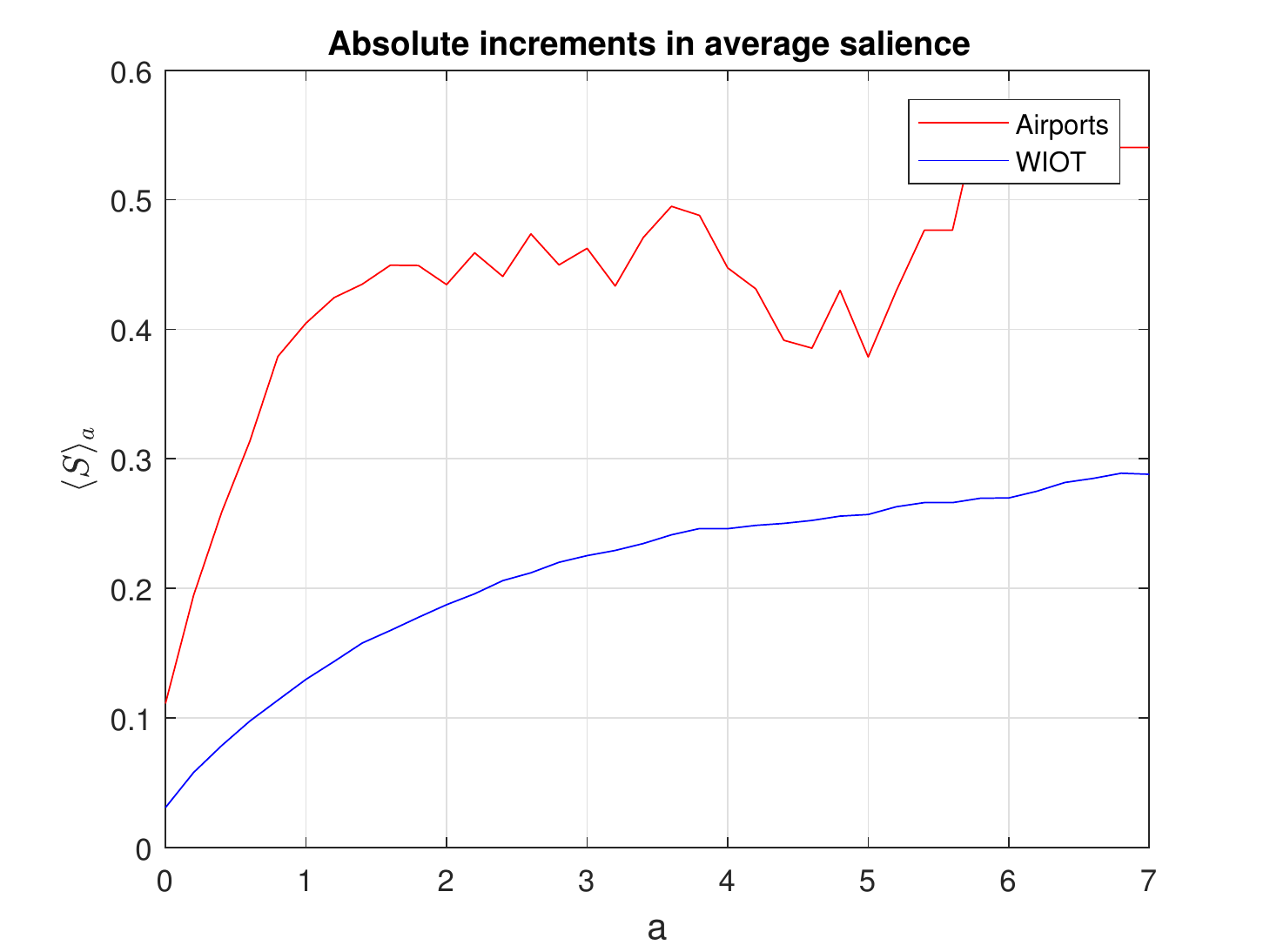}
\includegraphics[width=.45\linewidth]{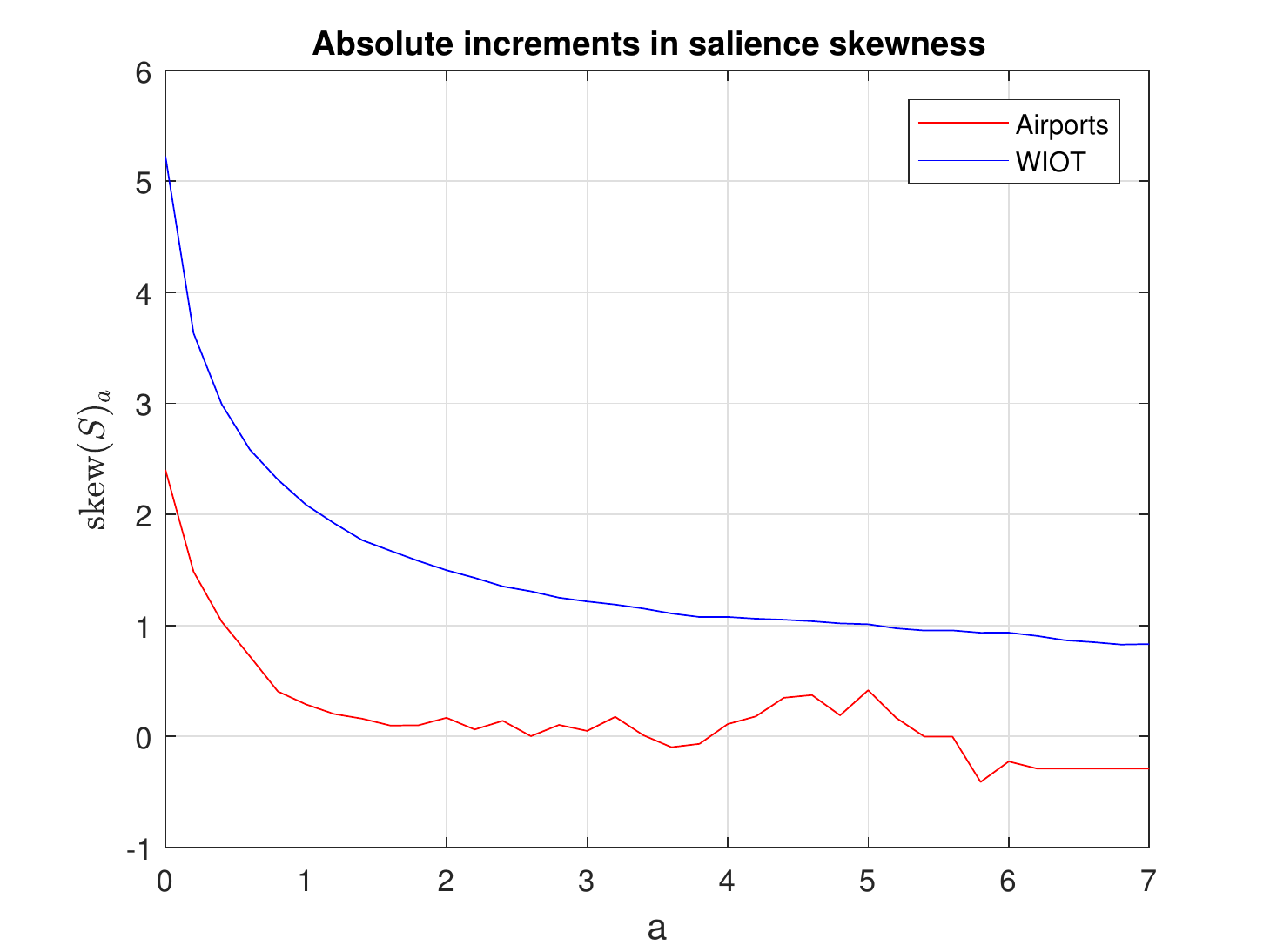}
\caption{\textbf{Left panel:} Average salience progressively calculated only in the links included in the backbones: $\; \langle S \rangle_a = \frac{1}{l_a} \sum_{(i,j) \in \mathcal{P}_a} S_{i j}$ where $l_a$ is the number of links in the backbone $\mathcal{P}_a$. \textbf{Right panel:} Skewness of the salience progressively calculated only in the links included in the backbones: $\;\textnormal{skew}(S)_a = \textnormal{skew}_{(i,j) \in \mathcal{P}_a} S_{i j}$ (S) where $l_a$ is the number of links in the backbone $\mathcal{P}_a$.}
\label{fig: salience}
\end{figure}

Interestingly, the P\'olya filter displays an empirical relationship with the salience. In both the WIOT and the US Airport network, we verify that, as we increase the parameter $a$, the filter has a tendency to retain links with higher salience. We show this in Fig. \ref{fig: salience} by plotting the mean and the skewness of the link salience distribution in both networks computed only in the links retained in the P\'olya backbones. As it can be seen, the mean increases (not necessary monotonically) while the skewness decreases as $a$ is raised.

The intuition behind this can be found once again in the ratio $r = k w / s$ (Eq. \eqref{eq:r_def}). Indeed, we have shown that links associated with a higher $r$ are typically assigned lower $p$-values by the P\'olya filter. The same can be said for the salience, whose scores appear to have a positive and statistically significant rank correlation with the corresponding values of $r$: $\textnormal{corr}(r,s) \approx 0.3$ in the US Airport network, and $\textnormal{corr}(r,s) \approx 0.2$ in the WIOT network.

\section{Additional comparisons between the P\'olya filter and other filtering techniques} \label{sec:sn8}

In this Section we present comparisons between the backbones generated by the P\'olya filter and those generated by other filtering techniques on two additional datasets. These are the Florida ecosystem \cite{ulanowicz2005network} and the High School network \cite{mastrandrea2015contact} (see Appendix \ref{sec:methods} for a description).

As done in the main text, we compare properties of the P\'olya backbones obtained at a certain level of statistical significance with those of the backbones obtained (at the same statistical significance) with other methods, i.e., the Hypergeometric Filter (HF) \cite{bonferroni_filter}, the Maximum-Likelihood filter (MLF) \cite{dianati2016unwinding}, the Enhanced Configuration Model (ECM) based on the canonical ensemble constrained both on degrees and strengths \cite{gemmetto2017irreducible}, the Noise-Corrected Bayesian filter (NC) proposed in \cite{coscia2017network}, and the Disparity Filter (DF) \cite{backbone_vespignani}, which in Appendix \ref{sec:sn3} we have shown to correspond to a large strength approximation of the P\'olya filter for $a=1$. For both the above datasets, we show comparisons across four main dimensions: the fraction of nodes retained in the backbone, the fraction of links retained, the salience-related optimality measure $O_1$ defined in Eq. \eqref{eq:optimality}, and the Jaccard similarity between the $B$ weights retained in the backbone and the top $B$ weights in the original network. 

\begin{figure}
\centering
\includegraphics[width=.45\linewidth]{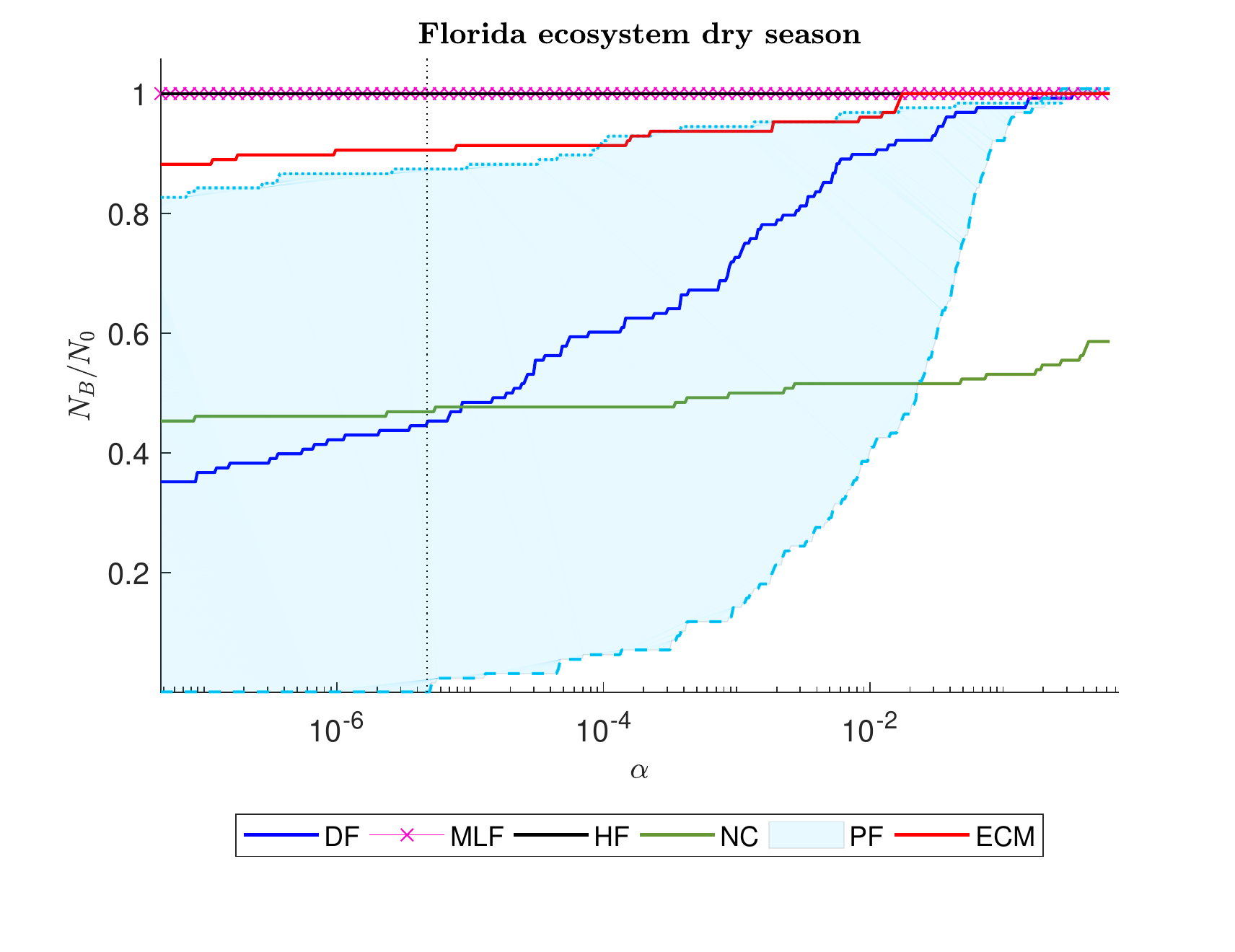}
\includegraphics[width=.45\linewidth]{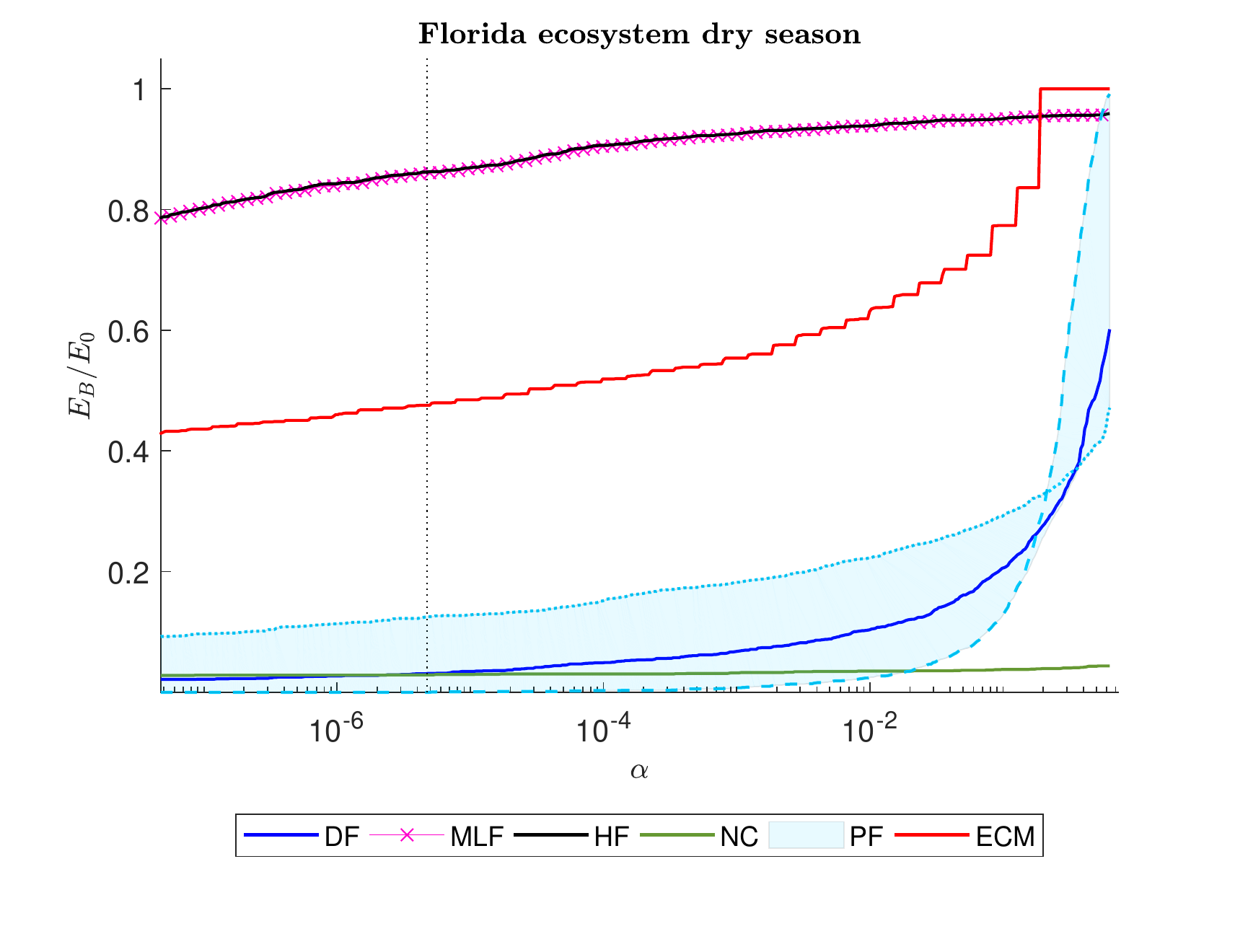}
\includegraphics[width=.45\linewidth]{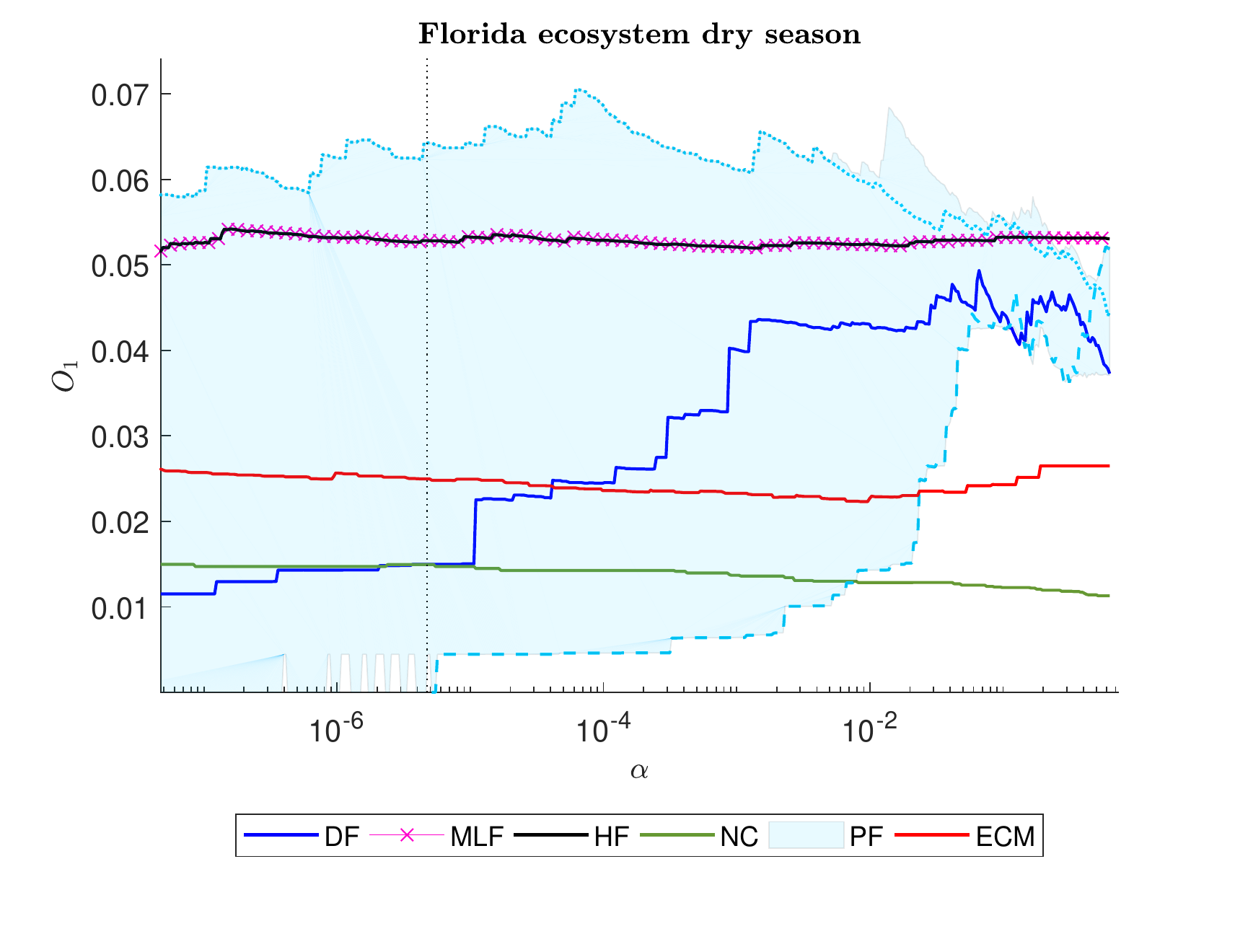}
\includegraphics[width=.45\linewidth]{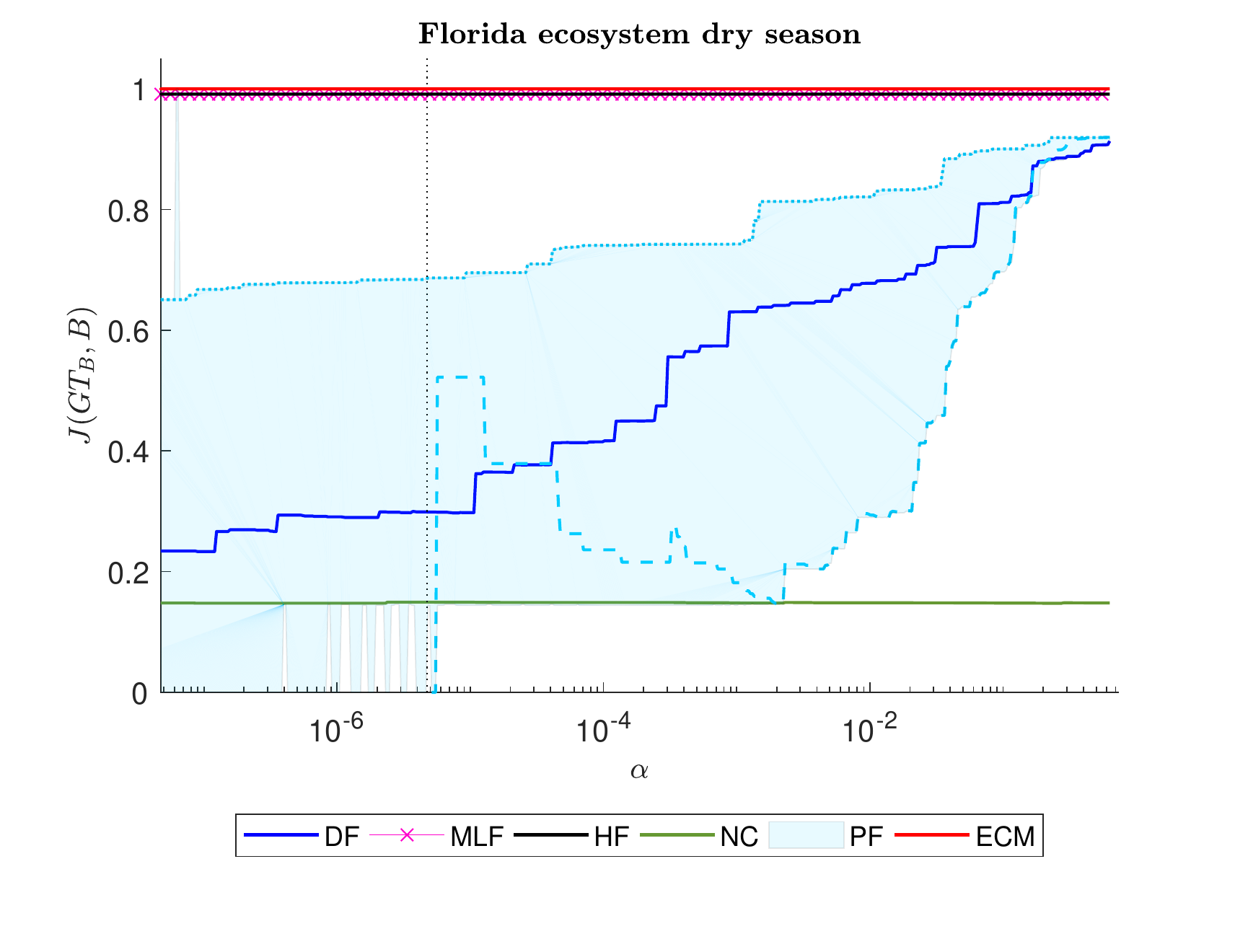}
\caption{Comparisons between the backbones generated for the Florida ecosystem network by the P\'olya filter (PF) and other network filtering methods, i.e., the Hypergeometric filter (HF), the Maximum-Likelihood filter (MLF), the Enhanced Configuration Model (ECM), the Noise-Corrected filter (NC), and the Disparity filer (DF). All quantities are shown as a function of the significance level used in the tests. TOP-LEFT: Fraction of nodes retained in the backbones. TOP-RIGHT: Fraction of edges retained in the backbones. BOTTOM-LEFT: Value of the salience-related measure $O_1$ defined in Eq. \eqref{eq:optimality}. BOTTOM-RIGHT: Jaccard similarity between the $B$ weights retained in the backbones and the top $B$ weights in the original networks. In all plots the light blue band correspond to all values measured in the P\'olya backbone families for $a \in [0.2,7]$, with the light blue solid (dashed) line corresponding to $a = 0.2$ ($a = 7$).}
\label{fig:Florida}
\end{figure}

\begin{figure}
\centering
\includegraphics[width=.45\linewidth]{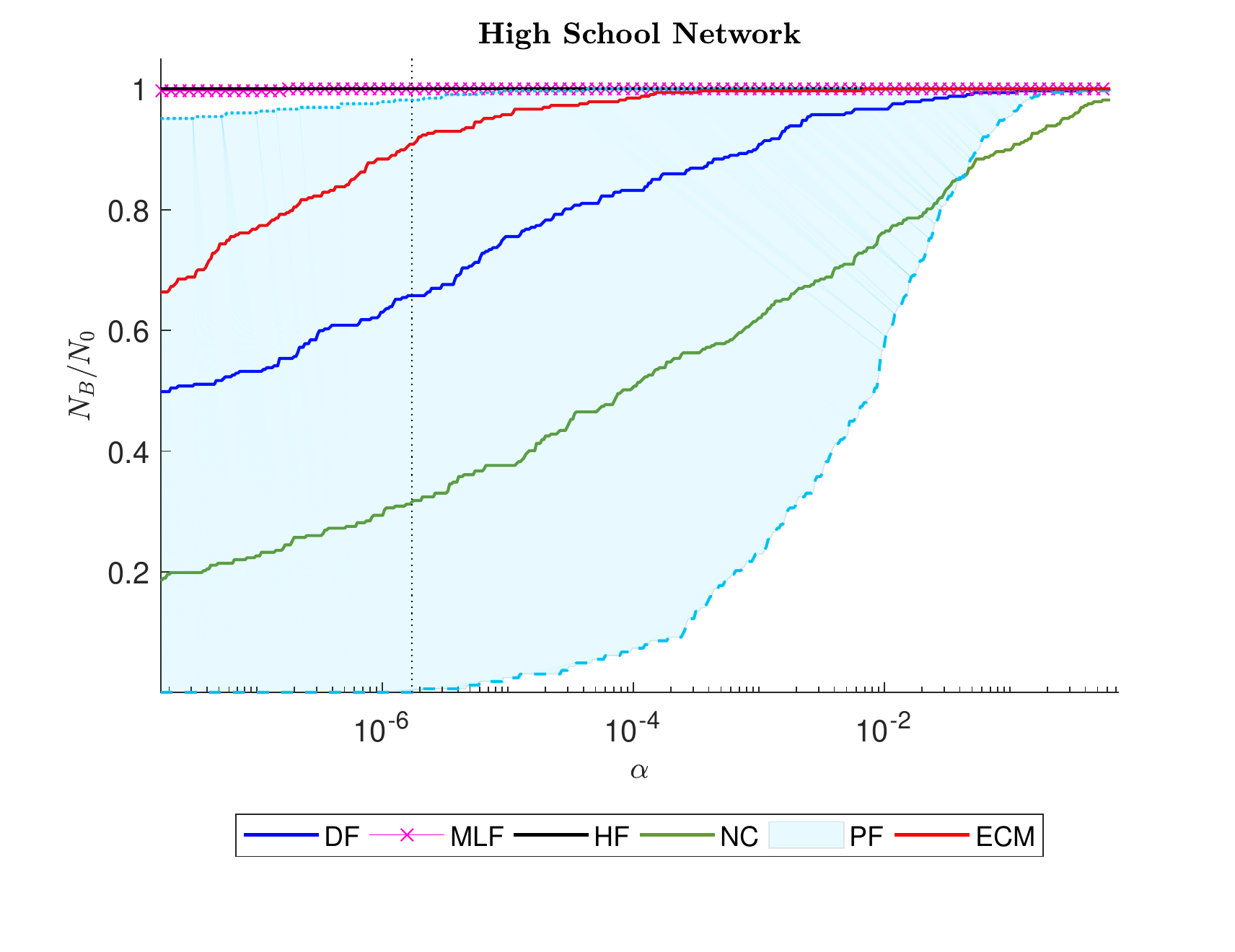}
\includegraphics[width=.45\linewidth]{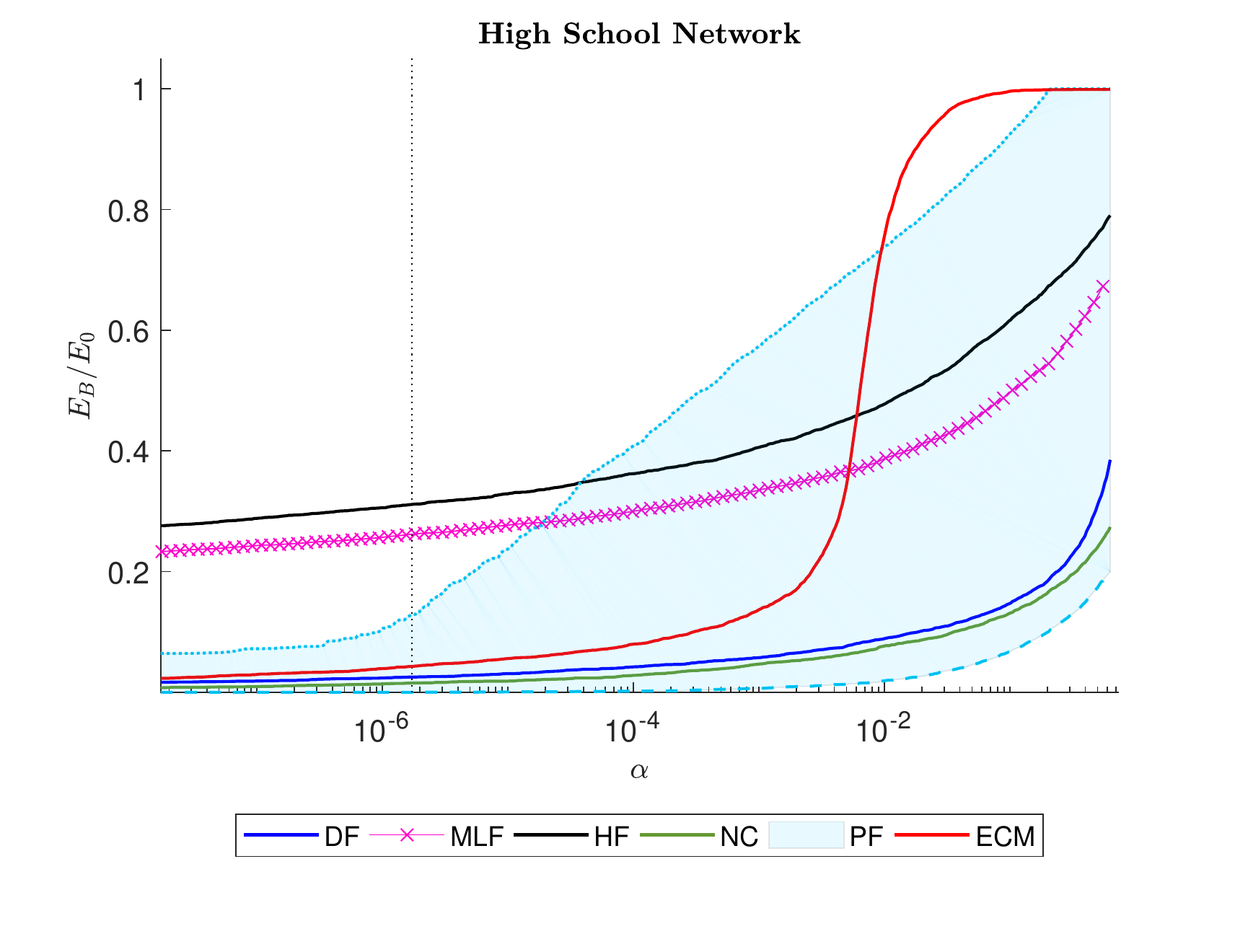}
\includegraphics[width=.45\linewidth]{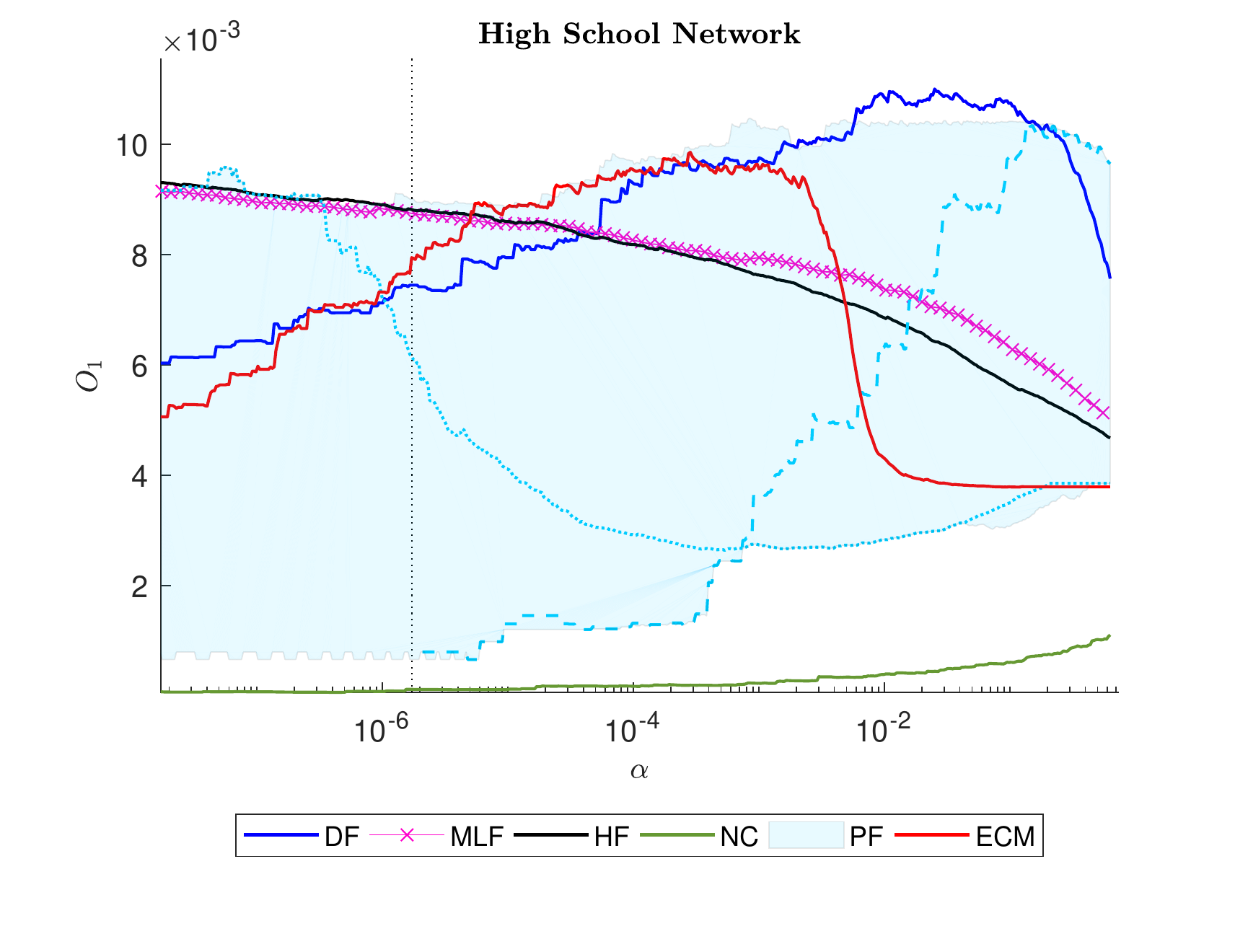}
\includegraphics[width=.45\linewidth]{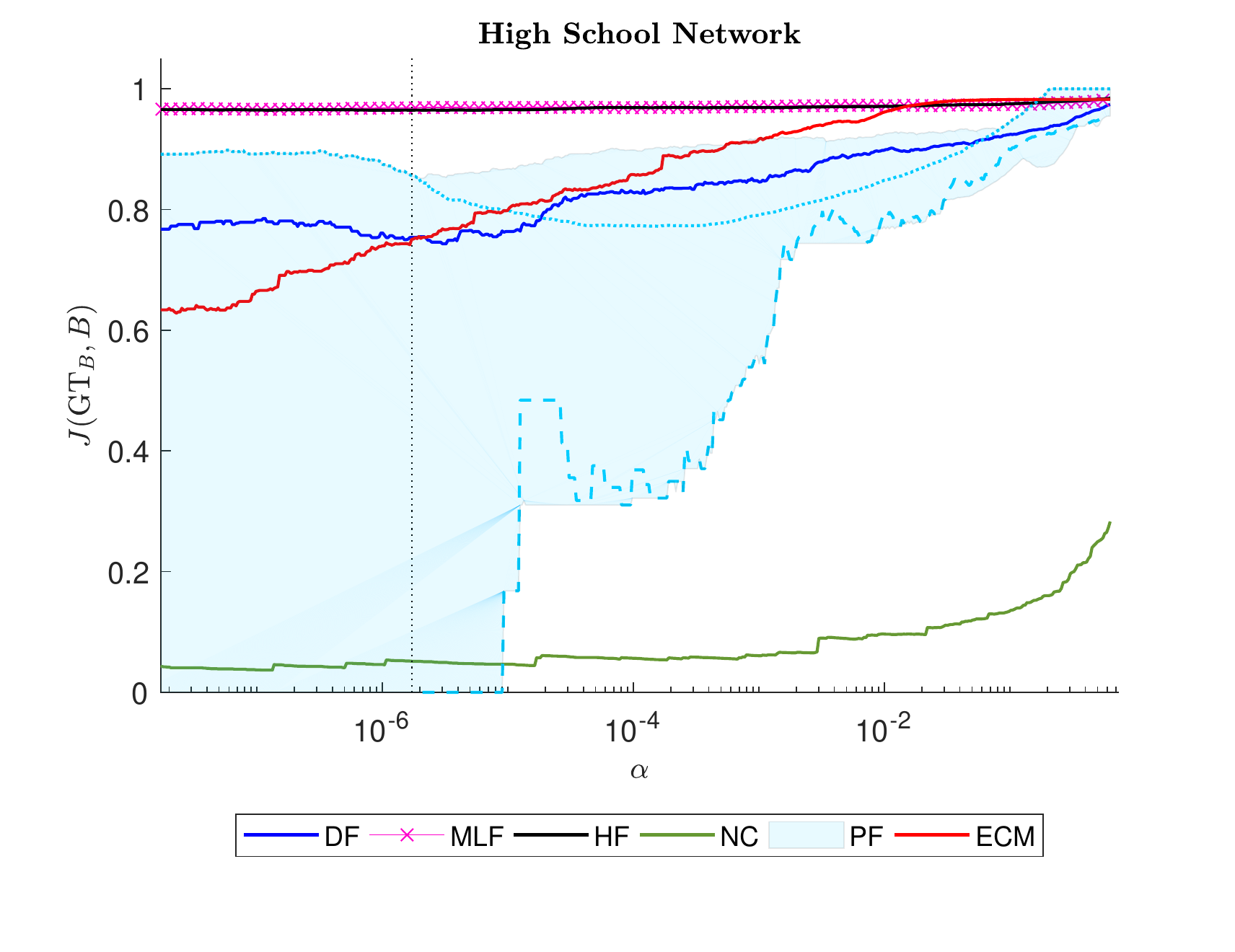}
\caption{Comparisons between the backbones generated for the High School network by the P\'olya filter (PF) and other network filtering methods, i.e., the Hypergeometric filter (HF), the Maximum-Likelihood filter (MLF), the Enhanced Configuration Model (ECM), the Noise-Corrected filter (NC), and the Disparity filer (DF). All quantities are shown as a function of the significance level used in the tests. TOP-LEFT: Fraction of nodes retained in the backbones. TOP-RIGHT: Fraction of edges retained in the backbones. BOTTOM-LEFT: Value of the salience-related measure $O_1$ defined in Eq. \eqref{eq:optimality}. BOTTOM-RIGHT: Jaccard similarity between the $B$ weights retained in the backbones and the top $B$ weights in the original networks. In all plots the light blue band correspond to all values measured in the P\'olya backbone families for $a \in [0.2,7]$, with the light blue solid (dashed) line corresponding to $a = 0.2$ ($a = 7$).}
\label{fig:HS}
\end{figure}

In Fig. \ref{fig:Florida} we report the results for the Florida network, while in Fig. \ref{fig:HS} we report results for the High School network. As in the main text, we see that P\'olya backbones are typically sparse, salient, and heterogeneous, and that the other methods we considered do not provide such combination. Indeed, the BF and MLF (whose results are extremely close along all dimensions), tend to preserve exceedingly high fractions of links. This was less evident in the examples shown in the main text (where two other methods ended up validating more links) but is apparent in the examples presented here, where both the BF and MLF validate almost all links in the Florida and HS networks, and do not filter out any node. This, obviously, translates into a very high Jaccard similarity between the weights in the backbone and the top weights in the original network, since almost none of these get filtered out.   

The NC method, on the other hand, provides the sparsest backbones of the methods we consider, and such backbones are heterogeneous as testified by the low Jaccard similarity between the weights on the links retained in them and the top links in the original networks. Yet, such links are not salient enough to compensate for such sparsity, as demonstrated by the very low values of the $O_1$ metrics achieved by the NC method. Hence, such a method provides parsimonious and non-trivial backbones, but it does so at the expense of salience, i.e., filtering out links that are globally important at the network-wide level. 

The ECM method represents an intermediate solution between the above. It provides rather parsimonious backbones, but it tends to do so simply by retaining the heaviest links in the network. This is particularly apparent in the case of the Florida network, where the $B$ links retained in the ECM backbone are exactly the heaviest $B$ links in the original network.

\begin{figure}
\centering
\includegraphics[width=.45\linewidth]{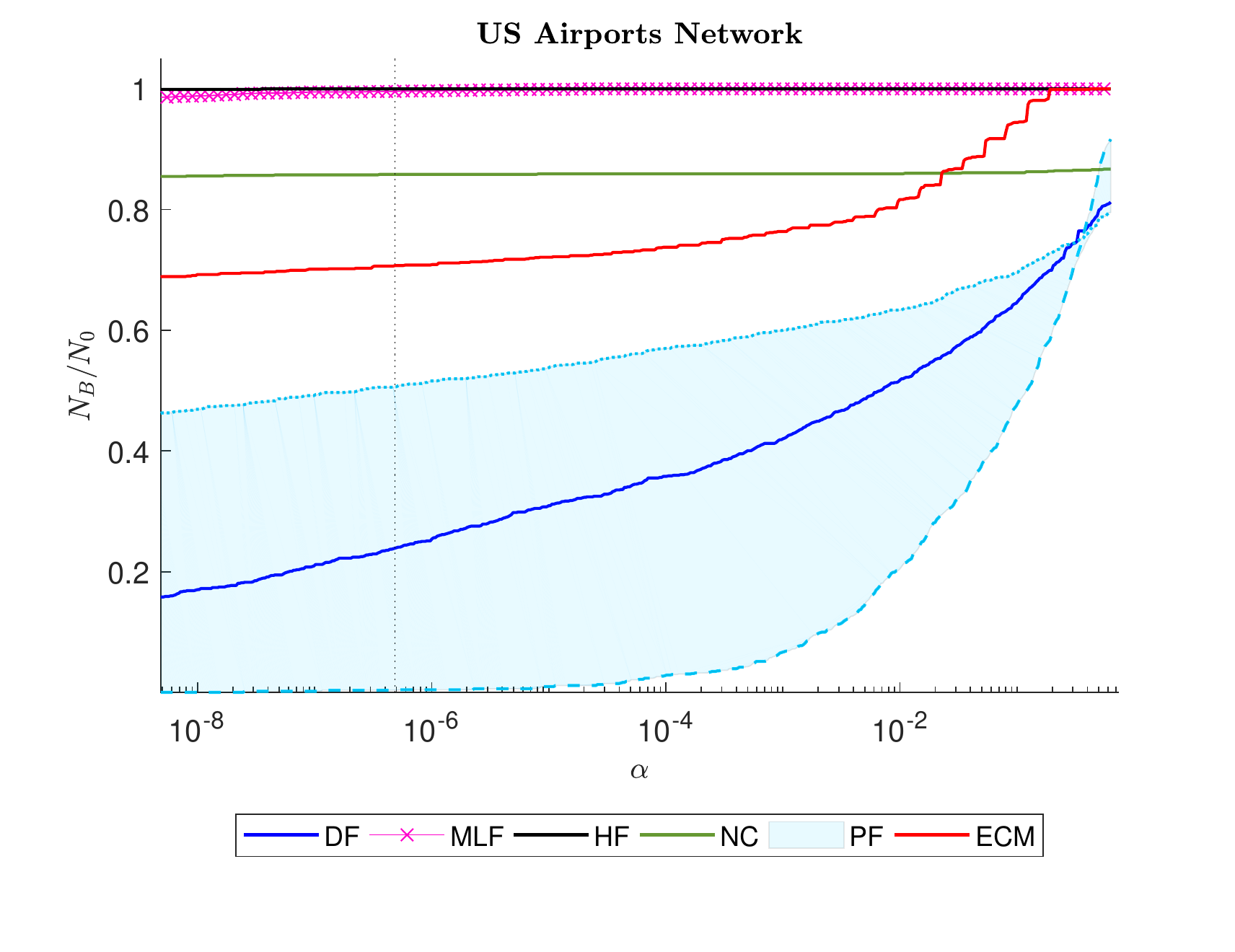}
\includegraphics[width=.45\linewidth]{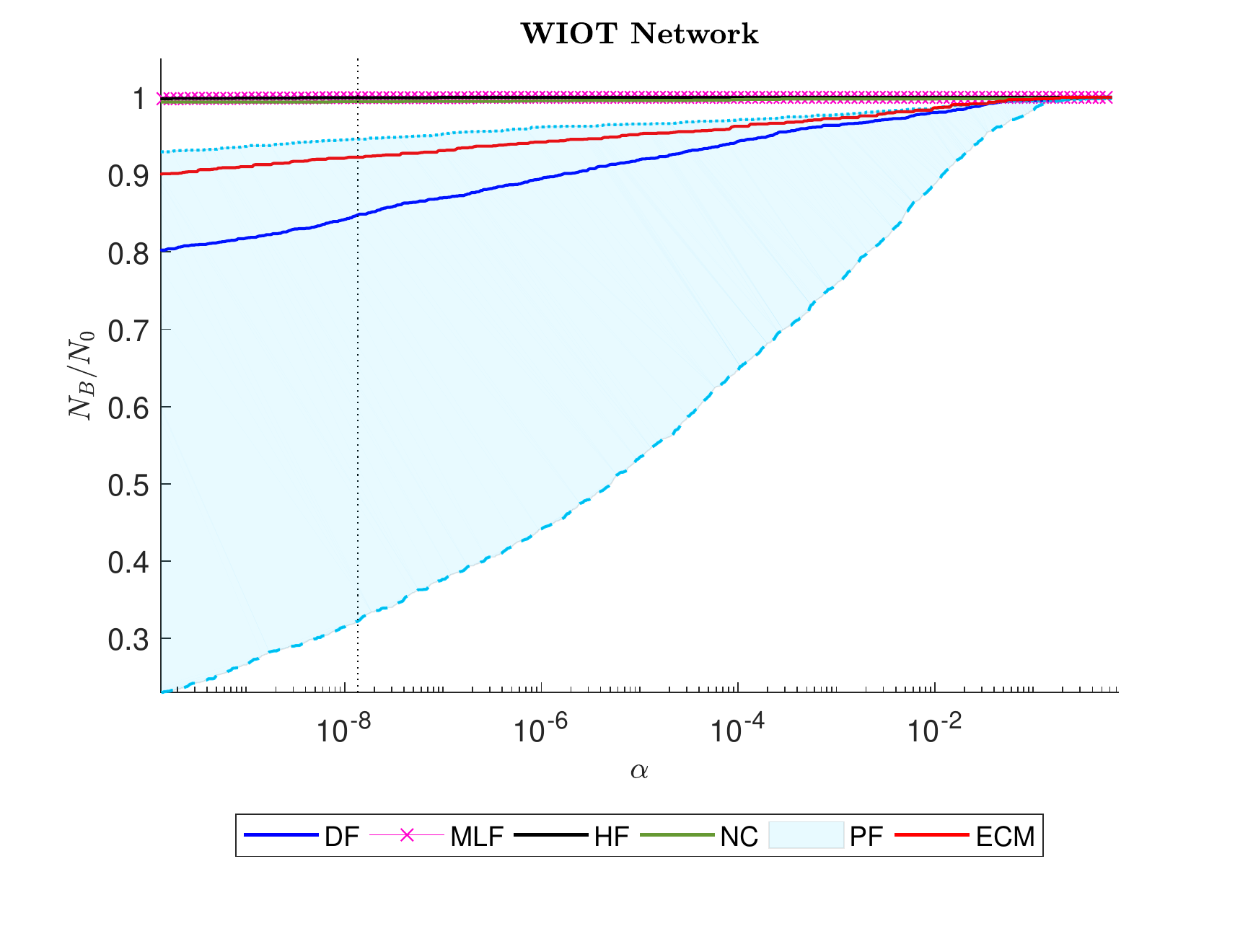}
\caption{Fraction of nodes retained in the P\'olya backbones of the US air transport (left) and WIOT (right) networks as a function of the significance level.}
\label{fig:air_WIOT}
\end{figure}

The DF corresponds to a large-strength approximation of the P\'olya filter for $a=1$ (see Appendix \ref{sec:sn3}). As such, it obviously occupies an intermediate position in the P\'olya family of backbones, and its effectiveness as a filtering tool largely depends on the specific network under study and its heterogeneity. Like other P\'olya backbones, it provides more parsimonious representations than other methods. On the other hand, the salience and heterogeneity of DF backbones vary significantly from network to network. For example, in the case of the Florida ecosystem network, the DF yields a heterogenous backbone (as testified by the low value of the Jaccard similarity measure) which, however, is not very salient compared to other P\'olya backbones. Conversely, in the case of the High School network, DF backbones are close to being optimal within the P\'olya family in terms of salience.

All in all, the above results reiterate the message of the main text, i.e., that the P\'olya filter main element of strength is its flexibility. Within reasonable ranges of the parameter $a$, all P\'olya backbones provide a parsimonious representation of the salient relationships in a network, while still retaining weights across multiple scales. Then, depending on the specific application or network, the parameter $a$ can be tuned to generate a backbone which is optimal with respect to a desired criterion.

For the sake of completeness, in Fig. \ref{fig:air_WIOT} we report the fraction of nodes retained in the backbones generated by the various methods we considered for the US air transport and WIOT networks. 

\section{Predicting trade volume in the WIOT network} \label{sec:sn10}

Following \cite{carvalho2014input,mcnerney2018production}, we propose a simple network-based regression model to predict changes in the trade volume between two nodes (representing industrial sectors) in the WIOT network. The model is as follows:
\begin{equation} \label{eq:linmodel}
\log(w_{i j}^{t+\tau}) = \beta_0 + \beta_1 A_{i j}^{t} + \beta_2 L_{i j}^{t} \ ,
\end{equation}
where 
\begin{itemize}
    \item $w_{ij}^{t}$ is the weight on the link between nodes $i$ and $j$ (i.e., the trade volume between the two corresponding industrial sectors) in year $t$.
    \item $A_{i j}^t$ is the element of the matrix $A^t_{i j} = w^t_{i j} / \sum_i w^t_{i j}$ in year $t$, i.e. the trade volume between nodes $i$ and $j$ normalized by the overall outgoing trade volume of node $j$.
    \item $L_{i j}^t$ is the year $t$ element of the Leontief matrix, defined as $L = (I-A^T)^{-1}$, where $A$ is defined above and $I$ is the identity matrix. The Leontief matrix is closely related to Katz centrality, and entry $L_{i j}^t$ quantifies the production required from sector $j$ in order to produce one unit of the good produced by sector $i$.
\end{itemize}

Note that the regression in Eq. \eqref{eq:linmodel} is defined only on links existing at time $t$ (i.e., $W_{i j}^t>0$). As mentioned in the main text, we calibrated the model over 5 years of data, from 2006 to 2010. We assume time-$t$ values in Eq. \eqref{eq:linmodel} to denote the values obtained after such calibration, and in the main text we show the results of the model's prediction for $\tau = 1,2,3$ (i.e., for the years from 2011 to 2013).

\ref{tab:regression} shows the results of the model's calibration when performed on the whole WIOT network, and on its P\'olya backbones for $a=1$ and $a=a_\mathrm{ML}=3.4$. As it can be seen, in all three cases the model's coefficients are highly significant, and the model as a whole is able to explain a good portion of the variance in data, as indicated by the $R^2$ coefficient. Notably, these increase when filtering the network, even though the number $N$ of links used to calibrate the model is reduced by more than two orders of magnitude when going from the full network to the $a=a_\mathrm{ML}$ P\'olya backbone. Also, upon filtering the network the importance of the weights, encoded in the matrix $A_{ij}^t$ and in its coefficient $\beta_1$ in Eq. \eqref{eq:linmodel}, decreases dramatically. Conversely, the importance of the Leontief matrix, quantified by its coefficient $\beta_2$, increases by roughly a factor $3$. This point is particularly significant, since the Leontief matrix is a non-local quantity which assesses the relevance of links from the viewpoint of the whole network they are embedded in. We interpret these results as a sign that P\'olya backbones, especially those obtained by tuning the filter to the network's specific heterogeneity, are highly informative, and contain links that are important both locally and globally.  

\begin{table}
\centering
  \caption{Regression table of the linear regression model in Eq. \eqref{eq:linmodel} calibrated on WIOT network data from 2006 to 2010. The three columns refer to the results obtained when calibrating the model on the full unfiltered network, and on its P\'olya backbones for $a=1$ and $a=a_\mathrm{ML}=3.4$.} 
  \label{tab:regression} 
{
\def\sym#1{\ifmmode^{#1}\else\(^{#1}\)\fi}
\begin{tabular}{l*{4}{c}}
\hline\hline\\[-1.8ex] 
            &\multicolumn{1}{c}{Unfiltered Networks}&\multicolumn{1}{c}{Backbones $\mathcal{P}_{a=1}$}&\multicolumn{1}{c}{Backbones $\mathcal{P}_{a=a_{ML}}$}\\
            &\multicolumn{1}{c}{(2006-2010)}&\multicolumn{1}{c}{(2006-2010)}&\multicolumn{1}{c}{(2006-2010)} \\
\hline\\[-2.2ex] 
$\beta_0$      &      1.61\sym{***}&      6.20\sym{***}&      7.12\sym{***}\\
            &    (0.00096)         &    (0.0090)         &     (0.017) \\
[1em]            
$\beta_1$     &     27.58\sym{***}&     4.52\sym{***}  &       3.21\sym{***}\\
            &   (0.043)         &   (0.064)         &    (0.079)          \\
$\beta_2$      &      0.018\sym{***}&      0.064\sym{***} &       0.058\sym{***} \\
            &   (0.00011)         &    (0.00073)         &     (0.00111)\\
[1em]

\hline\\[-2.2ex] 
$N$       &        2682840         &        48853         &        14784\\
$R^2=R^2_{adj}$   &       0.138         &       0.196         &  0.218\\
F statistic vs constant model &   $2.16 \times 10^{5}$ \sym{***} & $5.95 \times 10^{3}$ \sym{***} & $2.06 \times 10^{3}$ \sym{***} \\
\hline\hline\\[-1.8ex] 
\multicolumn{4}{l}{\footnotesize Standard errors in parentheses. Two-tailed test.}\\
\multicolumn{4}{l}{\footnotesize \sym{***} $p<0.0001$}\\
\end{tabular}
}
\end{table}

\bibliography{sample}

\end{document}